%% file: LCPartIf.tex
\documentclass[12pt,onecolumn]{IEEEtran}

\usepackage[colorlinks=false]{hyperref}
\usepackage[dvips]{graphicx}
\usepackage[usenames,dvipsnames]{color}
\usepackage{epsfig}
\usepackage{rotating}
\usepackage{amssymb}
\usepackage[boxed]{algorithm}
\usepackage[latin1]{inputenc}
\usepackage{verbatim}

\begin{document}

\input{custom1.tex}

\title{Generalised Bent Criteria for Boolean Functions (I)}
\author{Constanza Riera\thanks{C. Riera is with the Depto. de \'{A}lgebra,
Facultad de Matem\'{a}ticas, Universidad Complutense de Madrid,
Avda. Complutense s/n, 28040 Madrid, Spain. E-mail: \texttt{criera@mat.ucm.es}. Supported by the Spanish Government Grant AP2000-1365, and the Marie Curie Scholarship.},
Matthew G. Parker\thanks{M.G.Parker is with the
  Selmer Centre, Inst. for Informatikk, H{\o}yteknologisenteret i Bergen,
  University of Bergen, Bergen 5020, Norway. E-mail: \texttt{matthew@ii.uib.no}.
  Web: \texttt{http://www.ii.uib.no/\~{}matthew/}}}

\date{\today}
\maketitle

%\begin{tiny}
%LCPartIf.tex, 13.12.04
%\end{tiny}

\begin{abstract}
Generalisations of the bent property of a boolean function are presented,
by proposing
spectral analysis with respect to a well-chosen set of local unitary transforms.
Quadratic boolean functions are related
to simple graphs and it is shown that the orbit
generated by successive Local Complementations on a graph can be found within
the transform spectra under investigation. The flat
spectra of a quadratic boolean function are related to modified versions
of its associated adjacency matrix.
\end{abstract}

\section{Introduction}
It is often desirable that a boolean function, $p$, used for
cryptographic applications,
is be highly {\em{nonlinear}}, where nonlinearity is determined by
examining the spectrum of $p$ with respect to (w.r.t.) the
{\em{Walsh Hadamard Transform}} (WHT), and where the nonlinearity is
maximised for those functions that minimise the magnitude of the spectral
coefficients. To be precise, define the boolean function
of $n$ variables $p : \mbox{GF}(2)^n \rightarrow \mbox{GF}(2)$, and the WHT by the $2^n \times 2^n$
unitary matrix
$U = H \otimes H \ldots \otimes H = \bigotimes_{i=0}^{n-1} H$,
where the Walsh-Hadamard kernel
$H = \frac{1}{\sqrt{2}}\begin{tiny} \left ( \begin{array}{rr}
1 & 1 \\
1 & -1
\end{array} \right ) \end{tiny}$, '$\bigotimes$' indicates the tensor product of
matrices, and unitary means that $UU^{\dag} = I_n$, where '$\dag$' means
transpose-conjugate and $I_n$ is the $2^n \times 2^n$ identity matrix. We further
define a length $2^n$ vector,
$s = (s_{0\ldots00},s_{0\ldots 01},s_{0\ldots 11},\ldots,s_{1\ldots 11})$
such that
$s_{{\bf{i}}} = (-1)^{p({\bf{i}})}$, where ${\bf{i}} \in \mbox{GF}(2)^n$.
Then the Walsh-Hadamard spectrum of $p$
is given by the matrix-vector product $P = Us$, where $P$ is a vector
of $2^n$ real spectral coefficients, $P_{{\bf{k}}}$, where ${\bf{k}} \in \mbox{GF}(2)^n$. 

The spectral coefficient, $P_{{\bf{k}}}$, with maximum magnitude tells us
the minimum (Hamming) distance, $d$, of $p$ to
the set of affine boolean functions, where
$d = 2^{n-1} - 2^{\frac{n-2}{2}}|P_{\bf{k}}|$.
By Parseval's Theorem, the extremal case occurs when all
$P_{{\bf{k}}}$ have equal magnitude, in which case $p$ is said to have
a {\em{flat}} WHT spectra, and is
referred to as {\em{bent}}. If $p$ is bent, then it is as far away as it can be
from the affine functions \cite{Meier:NL}, which is a desirable cryptographic
design goal.
It is an open problem to classify all bent
boolean functions, although many results are known
\cite{Dill:DS,MacW:Cod,Car:NewB,Dob:Bent}.

In this paper, we extend the concept of a bent boolean
function to some {\em{Generalised Bent Criteria}} for a boolean function,
where we now require
that $p$ has flat spectra w.r.t. one or more transforms
from a specified set of unitary transforms. 
The set of transforms we choose is not arbitrary
but is motivated by the choice of unitary transforms that are typically used to
action a local basis change for a pure $n$-qubit quantum state. We here apply such
transforms to a $n$-variable boolean function, and examine the resultant
spectra accordingly. In particular we apply all possible
transforms formed from $n$-fold tensor products of the identity
$I = \begin{tiny} \left ( \begin{array}{rr}
1 & 0 \\
0 & 1
\end{array} \right ) \end{tiny}$,
the Walsh-Hadamard kernel, $H$,
and the Negahadamard kernel \cite{Par:Bent},
$N = \frac{1}{\sqrt{2}}\begin{tiny} \left ( \begin{array}{rr}
1 & i \\
1 & -i
\end{array} \right ) \end{tiny}$, where $i^2 = -1$.
We refer to this set of transforms as the
{\em{ $\{I,H,N\}^n$ transform set}}, i.e. where all transforms are of the form
$\{I,H,N\}^n=\bigotimes_{j \in {\bf{R_I}}} I_j \bigotimes_{j \in {\bf{R_H}}} H_j
\bigotimes_{j \in {\bf{R_N}}} N_j$,
where the sets ${\bf{R_I}}, {\bf{R_H}}$ and ${\bf{R_N}}$ partition
$\{0,\ldots,n-1\}$, and $H_j$, say, is short for
\begin{small}
$I \otimes I \otimes \ldots \otimes I \otimes H \otimes I \otimes \ldots \otimes I$
\end{small}, with $H$ in the $j^{th}$ position.
There are
$3^n$ such transforms which act on a boolean function of $n$ variables
to produce $3^n$ spectra, each spectrum of which comprises $2^n$ spectral
elements (complex numbers). By contrast, the WHT can be described as $\{H\}^n$,
which is a transform set of size one, where the single resultant output
spectrum comprises just $2^n$ spectral elements.

\subsection{The Quantum Context}
The choice of $I$, $H$, and $N$, is motivated by their
importance for the construction of {\em{Quantum Error-Correcting Codes}} (QECCs).
This is because $I$, $H$, and $N$ are generators of the {\em{Local Clifford Group}}
\cite{Cald:Qua,Klapp:Cliff1} which is defined to be the set of matrices
that {\em{stabilize}} the group of Pauli matrices
\begin{footnote}{
The Pauli matrices are $I$,
$\sigma_x = \begin{tiny}  \left ( \begin{array}{rr} 0 & 1 \\ 1 & 0 \end{array} \right ) \end{tiny}$,
$\sigma_z = \begin{tiny}  \left ( \begin{array}{rr} 1 & 0 \\ 0 & -1 \end{array} \right ) \end{tiny}$,
and $\sigma_y = i\sigma_x\sigma_z$.
}\end{footnote}
which, in turn, form a basis for the set of local errors that act on the quantum code.
This implies that the set of {\em{locally-equivalent}} $n$-qubit quantum states, that
occur as joint eigenspectra w.r.t. $\{I,H,N\}^n$,
are equally robust to quantum errors from the Pauli error set.
Stabilizer QECCs can also be interpreted
as additive codes over GF$(4)$ \cite{Cald:Qua}. 

To evaluate
the quantum {\em{entanglement}} of a pure $n$-qubit state one should really
examine the spectra w.r.t. the infinite set of $n$-fold tensor products of all
$2 \times 2$ unitary matrices \cite{Par:QE}. Those states which
minimise all spectral magnitudes w.r.t. this infinite transform set
are as far
away as possible from all generalised affine functions and can be
considered to be highly entangled as the probability of observing
(measuring) any specific
qubit configuration is as small as possible, in any local measurement basis.
However it is computationally intractable
to evaluate, to any reasonable approximation, this continuous local unitary
spectrum beyond about $n = 4$ qubits (although approximate results up to
$n = 6$ are given in \cite{Par:QE}).
Therefore we choose, in this paper, a well-spaced subset of spectral points,
as computed by the set of $\{I,H,N\}^n$ transforms, from which to ascertain
approximate entanglement measures. Complete
spectra for such a transform set can be computed up to about $n = 10$ qubits
using a standard desk-top computer,
although partial results for higher $n$ are possible if the $n$-qubit
quantum state is represented by, say, a quadratic boolean function over $n$ variables.

\subsection{The Graphical Context}
The graphical description of certain pure quantum states was
investigated by Parker and Rijmen \cite{Par:QE}.
They proposed partial entanglement
measures for such states and made observations
about a {\em{Local Unitary (LU) Equivalence}} between graphs
describing the states w.r.t. the tensor product of $2 \times 2$
local unitary transforms.
These graphs were interpreted as quadratic boolean functions and it was
noted that bipartite quadratic functions are LU-equivalent to indicators
for binary linear error-correcting
codes. It was further observed
that physical quantum graph arrays are relevant to the work of
\cite{Par:QE} and were already under investigation in the guise of
{\em{cluster states}}, by Raussendorf and Briegel \cite{Raus:QC,Brie:Ent}.
These clusters form the 'substrate' for measurement-driven quantum computation.

Measurement-driven quantum computation on a {\em{quantum factor graph}} has
been discussed by Parker \cite{Par:QFG}.
Independent work by Schlingemann and Werner \cite{Sch:QG}, Glynn
\cite{Glynn:Graph,Glynn:Tome}, and by
Grassl, Klappenecker, and Rotteler \cite{Gras:QG} proposed to describe
{\em{stabilizer}} Quantum Error-Correcting Codes (QECCs) using graphs and,
for QECCs of dimension zero, the associated
graphs can be referred to as {\em{graph states}}.
The graph states are
equivalent to the graphs described by \cite{Par:QE} and therefore
have a natural representation using quadratic boolean functions.

In \cite{Par:QE} it was observed
that the complete graph, star graph,
and generalised GHZ (Greenberger-Horne-Zeilinger) states are all LU-equivalent.
It turns out that
LU-equivalence for graph states can be characterised, graphically, via
the {\em{Vertex-Neighbour-Complement}} (VNC) transformation, which
was defined by Glynn, in the context of QECCs, in \cite{Glynn:Graph}
(definition 4.2) and also,
independently, by Hein, Eisert and Briegel \cite{Hein:GrEnt},
and also by Van Den Nest and De Moor \cite{VanD:Gr}.
VNC is another name for {\em{Local Complementation}} (LC), as investigated by
Bouchet \cite{Bou:Iso,Bou:Tree,Bou:Grph} in the context of {\em{isotropic systems}}.
By applying LC to a graph $G$ we obtain a graph $G'$, in which
case we
say that $G$ and $G'$ are {\em LC-equivalent}. Moreover, the set of all
LC-equivalent graphs form an {\em{LC-orbit}}.
LC-equivalence translates into
the natural equivalence between $\GF(4)$ additive codes that keeps the weight
distribution of the code invariant \cite{Cald:Qua}.
There has been recent renewed interest in Bouchet's work motivated,
in part, by the application of {\em{interlace graphs}} to the reconstruction
of DNA strings \cite{Arr:DNA,Arr:Int}. In particular, various {\em{interlace polynomials}}
have been defined \cite{Arr:Int,Aig:Int,Arr:Int1,Arr:Int2} which mirror
some of the quadratic results of part II of this paper \cite{RP:BCII}. We point out
links to this work in part II but defer a thorough exposition of
these links to the future.

\subsection{The Boolean Context}
Spectral analysis w.r.t. $\{I,H,N\}^n$ also has application to
the cryptanalysis of classical cryptographic systems
\cite{DanAPC}. In particular, for a block cipher it models attack scenarios
where one has full
read/write access to a subset of plaintext bits and access to all ciphertext
bits, (see \cite{DanAPC} for more details).
The analysis of spectra w.r.t.
$\{I,H,N\}^n$ tells us more
about $p$ than is provided by the spectrum w.r.t. the WHT;
for instance, identifying relatively high generalised linear biases for
$p$ \cite{Par:SB}.
In Part I of this paper our aim is to introduce these new generalised
bent criteria. In Part II \cite{RP:BCII} we enumerate the flat spectra w.r.t.
$\{I,H,N\}^n$ and its subsets.
We are trying to answer the question:
which boolean functions are as far
away as possible from the set of generalised affine functions as defined by the
rows of $\{I,H,N\}^n$?
\begin{footnote}{
A row of $U_0 \otimes U_1 \otimes \ldots \otimes U_{n-1}$ for $U_i$ a $2 \times 2$
unitary matrix can always be written as
$u = (a_0,b_0) \otimes (a_1,b_1) \otimes \ldots \otimes (a_{n-1},b_{n-1})$, where $a_i,b_i$
are complex numbers. For $\al$ a $r^{th}$ complex root of 1, and $m$ an integer modulus,
we can approximate an unnormalised version of $u$ by
$u \simeq m({\bf{x}})\al^{p({\bf{x}})}$, for some appropriate choice of integers $s$ and $r$,
where $m : \mbox{GF}(2)^n \rightarrow \mbox{GF}(s)$,
$p : \mbox{GF}(2)^n \rightarrow \mbox{GF}(r)$, and ${\bf{x}} \in \mbox{GF}(2)^n$, such that
the ${\bf{j}}^{th}$ element of $u$,
$u_{\bf{j}} = m({\bf{j}})\al^{p({\bf{j}})}$, where ${\bf{j}} \in \mbox{GF}(2)^n$ and
$u_{\bf{j}}$ is interpreted as
a complex number. When
$u$ is fully-factorised using the tensor product then $m$ and $p$ are affine functions and
we say that $u$ represents a generalised affine function (see \cite{Par:QE}, Section 5, for more details).
}\end{footnote}

The classification of bent quadratic (degree-two) boolean functions
is well-known \cite{MacW:Cod}, and is facilitated because the bent criteria
is an
invariant of affine transformation of the input variables. However,
the classification of generalised bent criteria for a quadratic
boolean function w.r.t. the $\{I,H,N\}^n$ transform set is new, and the
generalised
bent criteria are not, in general, invariant to affine transformation of the
inputs. This paper characterises these generalised bent criteria
for both quadratic and more general boolean functions. We associate a quadratic boolean
function with an undirected graph, which allows us to interpret spectral flatness
with respect to $\{I,H,N\}^n$ as a {\em{maximum rank}} property of suitably
modified adjacency matrices.
We interpret LC as an operation on quadratic boolean functions,
and as an
operation on the associated adjacency matrix, and we also identify the
LC-orbit with
a subset of the flat spectra w.r.t. $\{I,H,N\}^n$.
The spectra w.r.t. $\{I,H,N\}^n$ motivates us to examine
the properties of the WHT
of all ${\mathbb{Z}}_4$-linear offsets of boolean functions, the WHT of
all subspaces of boolean functions that can be obtained by fixing a subset
of the variables, the
WHT of all ${\mathbb{Z}}_4$-linear offsets of all of the above subspace boolean functions,
the WHT of each member
of the LC-orbit, and the distance of boolean functions to all
${\mathbb{Z}}_4$-linear functions. This leads us to
prove the following:
\cb
     All quadratic boolean functions are {\em{bent$_4$}}, {\em{I-bent}} and {\em{I-bent$_4$}}. \\
     Not all quadratic boolean functions are {\em{LC-bent}}. \\
     All boolean functions are {\em{I-bent$_4$}}. \\
     Not all boolean functions are {\em{bent$_4$}} or {\em{I-bent}}. \\
     There are no {\em{${\mathbb{Z}}_4$-bent}} or {\em{Completely I-bent$_4$}} boolean functions.
\ce
where the above terms for generalised bent criteria will be made clear in the sequel.
We are able to characterise
and analyse the criteria for quadratic boolean functions by
considering properties of the adjacency
matrix for the associated graph state.

\subsection{Paper Overview}
For the interested reader, Appendix A reviews the graph state and its
intepretations in the literature.
%as a zero-dimension QECC, as a self-dual additive code over GF$(4)$,
%as a self-dual additive code over ${\mathbb{Z}}_4$
%as an isotropic system, and as a quadratic boolean function.
%Finally we note that
%bipartite graphs have an interpretation as binary linear error-correcting codes which
%is made explicit via the quadratic boolean function representation.
In Section \ref{LCGraph} we review LC as an operation on an
undirected graph \cite{Glynn:Graph,Glynn:Tome}, and provide an
algorithm for LC in terms of the adjacency matrix of the graph.
In Section \ref{LCLUT}, we show that the LC-orbit of a quadratic
boolean function lies within the set of
transform spectra w.r.t. tensor products of the
$2 \times 2$ matrices, $I$, $\sqrt{-i\sigma_x}$, and
$\sqrt{i\sigma_z}$, where $\sigma_x$ and $\sigma_z$ are Pauli matrices.
We also show, equivalently, that the orbit lies within the
spectra w.r.t. $\{I,H,N\}^n$.
We show that doing LC to vertex $x_v$ can be
realised by the application of the Negahadamard kernel, $N$, to position 
$v$ (and the identity matrix to all other positions) of the bipolar
vector $(-1)^{p({\bf{x}})}$, i.e.
$$ \omega^{4p'({\bf{x}}) + a({\bf{x}})}(-1)^{p'({\bf{x}})} = U_v(-1)^{p({\bf{x}})} =
I \otimes \cdots \otimes I \otimes N \otimes I \otimes \cdots \otimes
I \ (-1)^{p({\bf{x}})} \enspace,$$ 
where $p({\bf{x}})$ and $p'({\bf{x}})$ are quadratic, $p'({\bf{x}})$
is obtained by applying LC to variable $x_v$,
$\omega = \sqrt{i}$, and $a({\bf{x}})$ is any offset over ${\mathbb{Z}}_8$.
In Appendix B we identify spectral symmetries that hold for $p({\bf{x}})$ of any
degree w.r.t. $\{I,H,N\}^n$.
In Section \ref{BentSection}, we introduce the concepts of {\it bent$_4$},
{\it ${\mathbb{Z}}_4$-bent}, {\it (Completely) I-bent},
{\it LC-bent}, and {\it (Completely) I-bent$_4$}
boolean functions, and show how, for quadratic boolean
functions, these properties can be evaluated by examining the ranks of
suitably modified versions of the adjacency matrix.

\section{Local Complementation (LC)}
\label{LCGraph}
Given a graph $G$ with adjacency matrix $\Gamma$, define
its {\em complement} to be the graph with adjacency matrix $\Gamma+I + {\bf 1}\ (\mo 2)$,
where $I$ is the identity matrix and ${\bf 1}$ is the all-ones matrix.
Let ${\cal{N}}(v)$ be the set of neighbours of vertex, $v$, in the graph, $G$, i.e.
the set of vertices connected to $v$ in $G$.

\begin{df} Define the action of {\em LC} (or {\em vertex-neighbour-complement (VNC)}) on a graph $G$ at vertex $v$ as
the graph transformation obtained by replacing the subgraph $G[{\cal{N}}(v)]$
by its complement.
\end{df}

By Glynn (see \cite{Glynn:Graph}), a self-dual quantum code $[[n,0,d]]$
corresponds to a graph on $n$ vertices, which may be assumed to be connected
if the code is indecomposable. It is shown there that two graphs $G$ and $H$
give equivalent self-dual quantum codes if and only if $H$ and $G$ are LC-equivalent.

For a study of the group of compositions of local complementations, see
\cite{Bou:Iso,Bou:Grph,Bou:Tree,Cou:Ver}, which describe the relation between local
complementation and {\em isotropic systems}. Essentially, a suitably-specified
isotropic system has graph presentations $G$ and $G'$ iff $G$ and $G'$ are
locally equivalent w.r.t. local complementation.

\subsection{LC in terms of the adjacency matrix}
Let $p({\bf{x}}) : F_2^n \rightarrow F_2$ be a (homogeneous) quadratic boolean
function, defined by,
$$ p({\bf{x}}) = \sum_{0\leq i < j\leq n-1} a_{ij}x^ix^j \enspace.$$
We can express $p({\bf{x}})$ by the adjacency matrix of its
associated graph,
$\Gamma$, such that $\Gamma(i,j) = \Gamma(j,i) = a_{ij}$, $i < j$,
$\Gamma(i,i) = 0$.
The LC operation on the graph associated to $p({\bf{x}})$ can be expressed
in terms of the adjacency
 matrix. Without loss of generality, we show how the matrix changes from
 $\Gamma$ to $\Gamma_0$
after doing LC on vertex $x_{0}$:

\begin{small}
$$\Gamma_{0}=\left(\begin{array}{cccccc}
0 & a_{01} & a_{02} & a_{03} & \ldots & a_{0n}\\
a_{01} & 0 & a_{12}+a_{01}a_{02} &  a_{13}+a_{01}a_{03} & \ldots & a_{1n}+a_{01}a_{0,n-1}\\
a_{02} & a_{12}+a_{01}a_{02} & 0 & a_{23}+a_{02}a_{03} & \ldots & a_{2n}+a_{02}a_{0,n-1}\\
a_{03} & a_{13}+a_{01}a_{03} & a_{23}+a_{02}a_{03} & 0 & \ldots & a_{3n}+a_{03}a_{0,n-1}\\
\vdots & \vdots & \vdots & \vdots & \ddots & \vdots\\
a_{0,n-1} & a_{1,n-1}+a_{01}a_{0,n-1} & a_{2,n-1}+a_{02}a_{0,n-1} & a_{3,n-1}+a_{03}a_{0,n-1} & \ldots & 0\end{array}\right) \enspace.$$
\end{small}

The general algorithm, mod 2, is
$$\left\{\begin{array}{l}
\Gamma_{v}(i,j)=\Gamma(i,j)+\Gamma(v,i)*\Gamma(v,j),\ \  i<j,\ \ \  i,j=1,\ldots,n\\
\Gamma_{v}(i,i)=0\ \ \forall i\\
\Gamma_{v}(j,i)=\Gamma_{v}(i,j),\ \ \  i>j 
\end{array}\right.$$
where $\Gamma_{v}$ is the adjacency matrix of the function after doing LC to
the vertex $x_{v}$.
%\begin{proof}
%We look at the ANF of the function.
%Applying LC to vertex $x_{v}$ only affects the relations between its neighbours.
% That is, it changes the relations of $x_{i}$ 
%if and only if this vertex is a neighbour of $x_v$ (i.e., if and only if
%$a_{vi}=a_{iv}=1$). We thus have two cases:
%$$\begin{array}{l}
%1)\ a_{vi}=0\Rightarrow a_{ij}^{v}=a_{ij}\ \ \forall j=1,\ldots,n, \m{ where } a_{ij}=\Gamma(i,j),\ \  a_{ij}^{v}=\Gamma_{v}(i,j)\\
%2)\ a_{vi}=1\Rightarrow a_{ij}^{v}=a_{ij}\ \m{ if } a_{vj}=0;\ a_{ij}^{v}=a_{ij}+1\ \m{ if } \left\{ \begin{array}{ll}
%a_{ij}^{v}=a_{ij} & \textrm{if}\  a_{vj}=0\\
%a_{ij}^{v}=a_{ij}+1 & \textrm{if}\  a_{vj}=1\end{array}\right.
%\end{array}$$

%Finally, we can write this in a neater way, saying that 
%$a_{ij}^{v}=a_{ij}+a_{vi}a_{vj}$. We obtain the deduced algorithm by writing
%the above as a matrix.
%\end{proof}

\section{Local Complementation (LC) and Local Unitary (LU) Equivalence}
\label{LCLUT}
Hein et al \cite{Hein:GrEnt} state that LC-Equivalence (and therefore
Local Unitary (LU) Equivalence)
of graph states is obtained via successive transformations of the form,
\beg U_v(G) = (-i\sigma_x^{(v)})^{1/2} \prod_{b \in {\cal{N}}_v} (i\sigma_z^{(b)})^{1/2} \label{LUEquiv} \enspace,\eeg
where
$\sigma_x = \begin{tiny}  \left ( \begin{array}{rr} 0 & 1 \\ 1 & 0 \end{array} \right ) \end{tiny}$
and
$\sigma_z = \begin{tiny}  \left ( \begin{array}{rr} 1 & 0 \\ 0 & -1 \end{array} \right ) \end{tiny}$
are Pauli matrices,
the superscript $(v)$ indicates that the Pauli matrix acts on qubit $v$ (with
$I$ acting on all other qubits)
\begin{footnote}{For instance, $\sigma_x^{(2)} =
I \otimes I \otimes \sigma_x \otimes I \otimes \ldots \otimes I$.
} \end{footnote}, and ${\cal{N}}_v$
comprises the neighbours of qubit $v$ in the graphical representation.
Define matrices $x$ and $z$ as follows:
$$ x = (-i\sigma_x)^{1/2} = \frac{1}{\sqrt{2}}
\begin{small} \left ( \begin{array}{rr} -1 & i \\ i & -1 \end{array} \right ) \end{small} $$
and
$$ z = (i\sigma_z)^{1/2} =
\begin{small} \left ( \begin{array}{rr} w & 0 \\ 0 & w^3 \end{array} \right ) \end{small} \enspace,$$
where $w = e^{2\pi i/8}$.
Furthermore, let
$I = \begin{tiny}  \left ( \begin{array}{rr} 1 & 0 \\ 0 & 1 \end{array} \right ) \end{tiny}$.

Define ${\bf{D}}$ to be
the set of $2 \times 2$ diagonal or anti-diagonal local unitary matrices,
i.e. of the form
$\begin{tiny}  \left ( \begin{array}{rr} a & 0 \\ 0 & b \end{array} \right ) \end{tiny}$
or
$\begin{tiny}  \left ( \begin{array}{rr} 0 & a \\ b & 0 \end{array} \right ) \end{tiny}$,
for some $a$ and $b$ in $\mathbb{C}$. We make extensive use of the fact that a final multiplication
of a spectral vector by tensor products of members of ${\bf{D}}$ does not change spectral
coefficient magnitudes. In this sense a final multiplication by tensor products of members of ${\bf{D}}$
has no effect on the final spectrum and does not alter the underlying graphical
interpretation.
For instance, applying $x$ twice to the same qubit is the same as applying
$x^2 = \begin{tiny}  \left ( \begin{array}{rr} 0 & -i \\ -i & 0 \end{array} \right ) \end{tiny}$,
which is in ${\bf{D}}$.
Therefore we can equate $x^2$ with the
identity matrix, i.e.
$x^2 \simeq I = \begin{tiny}  \left ( \begin{array}{rr} 1 & 0 \\ 0 & 1 \end{array} \right ) \end{tiny}$.
Similarly, the action of any $2\times 2$ matrix from ${\bf{D}}$ on a specific
vertex is 'equivalent' the action of the identity on the same vertex.
Note that $z \in {\bf{D}}$. The same equivalence holds over $n$ vertices, so
we define an equivalence relation with respect to a tensor product of members
of ${\bf{D}}$ by the symbol '$\simeq$'. 
\begin{df} Let $u$ and $v$ be two $2 \times 2$ unitary matrices. Then,
$$ u \simeq v \mv \Leftrightarrow \mv u = dv, \mf d \in {\bf{D}} \enspace.$$\label{equiv} \end{df}
This equivalence relation allows us to simplify the concatenation of actions of $x$ and $z$ on a specific qubit.

{\bf{Remark:}} Note that $u \simeq v$ cannot be deduced from (and does not imply) $u=vd$ for some
$d \in {\bf{D}}$.

We now show that the LC-orbit of an $n$-node graph is found as a subset of
the transform spectra w.r.t. $\{I,x,xz\}^n$.
Subsequently, it
will be shown that we can alternatively find the LC-orbit as a subset of the
transform set w.r.t. $\{I,H,N\}^n$.
We then re-derive the single LC operation on a graph from the
application of $x$ (or $N$) on a single vertex.
\subsection{The LC-orbit Occurs Within the $\{I,x,xz\}^n$ Set of Transform Spectra}
\label{LCMultispectra}
We summarise the result of (\ref{LUEquiv}) as follows.
\begin{lem}
Given graphs $G$ and $G'$ as represented by the quadratic boolean functions,
$p({\bf{x}})$ and $p'({\bf{x}})$, then $G$ and $G'$ are in the same LC-orbit
iff $(-1)^{p'({\bf{x}})} \simeq
U_{v_{t-1}}U_{v_{t-2}}\ldots U_{v_{0}}(-1)^{p({\bf{x}})}$
for some series of $t$ local unitary transformations, $U_{v_i}$.
\label{GEq}
\end{lem}
From Lemma \ref{GEq} we see that,
by applying $U_v(G)$ successively for various $v$ to an initial state, one can
generate all
LC-equivalent graphs within a finite number of steps.
(It is evident from the action of LC on a graph that any LC-orbit
must be of finite size). Instead of applying $U$
successively, it would be nice to identify a (smaller) transform set in which all
LC-equivalent graphs exist as the spectra, to within a post-multiplication
by the tensor product of matrices from ${\bf{D}}$.
One can deduce from definition \ref{equiv} that $zx \simeq x$,
and it is easy to verify that
\begin{lem}  $zxx \simeq I$, and $xzx \simeq zxz$ \label{lem1} \end{lem}
With these definitions and observations we can derive the following theorem.
\begin{thm}
To within subsequent transformation by tensor
products of matrices from ${\bf{D}}$, the LC-orbit of the graph, $G$, over $n$ qubits occurs
within the spectra of all possible tensor product
combinations of the $2 \times 2$ matrices, $I$, $x$, and $xz$.
There are $3^n$ such transform spectra.
\label{thm1}
\end{thm}
\begin{proof}
For each vertex in $G$, consider every possible product of the two matrices, $x$, and $z$.
Using the equivalence relationship and lemma \ref{lem1},

\cb
\begin{tabular}{lll}
$xxx \simeq x$ & \mf & $zxx \simeq I$ \\
$xxz \simeq I$ & \mf &  $zxz \simeq xz$ \\
$xzx \simeq zxz \simeq xz$ & \mf &  $zzx \simeq x$ \\
$xzz \simeq zxzz \simeq xzxz \simeq xxzx \simeq x$ & \mf &  $zzz \simeq I \enspace.$ \\
\end{tabular}
\ce

%\begin{tabular}{ll}
%\bite
%        \item $xxx \simeq x$
%        \item $xxz \simeq I$
%        \item $xzx \simeq zxz \simeq xz$
%        \item $xzz \simeq zxzz \simeq xzxz \simeq xxzx \simeq x$
%\eite
%&
%\bite
%        \item $zxx \simeq I$
%        \item $zxz \simeq xz$
%        \item $zzx \simeq x$
%        \item $zzz \simeq I \enspace.$
%\eite
%\end{tabular}
Thus, any product of three or more instances of $x$ and/or $z$ can always be reduced to
$I$, $x$, or $xz$. Theorem \ref{thm1} follows by recursive application of (\ref{LUEquiv}) with these rules, and
by noting that the rules are unaffected by the tensor product expansion over $n$ vertices.
\end{proof}
For instance, for $n = 2$, the LC-orbit of the graph represented by
the quadratic function $p({\bf{x}})$ is found as a subset of 
the $3^2 = 9$ transform spectra of $(-1)^{p({\bf{x}})}$ w.r.t. the transforms
$I \otimes I$, $I \otimes x$, $I \otimes xz$, $x \otimes I$,
$x \otimes x$, $x \otimes xz$, $xz \otimes I$, $xz \otimes x$, and
$xz \otimes xz$.
Theorem \ref{thm1} gives a trivial and very loose upper bound on the maximum size
of any LC-orbit over $n$ qubits, this bound being $3^n$. It has been
computed in \cite{DanQECC} that the number of LC-orbits for connected graphs for
$n = 1$ to $n = 12$ are
$1,1,1,2,4,11,26,101,440,3132,40457$, and $1274068$, respectively
(see also \cite{Hein:GrEnt,Glynn:Tome,Hohn:Klein,Dan:Dat,Slo:Seq}).
\subsection{The LC-orbit Occurs Within the $\{I,H,N\}^n$ Set of Transform Spectra}
\label{IHNsection}
One can verify that $N \simeq x$ and $H \simeq xz$. Therefore
one can replace $x$ and $xz$ with $N$ and $H$, respectively, so the transform
set, $\{I,xz,x\}$ becomes $\{I,H,N\}$.
This is of theoretical interest because $H$ defines a 2-point (periodic) Discrete Fourier
Transform matrix, and $N$ defines a 2-point negaperiodic Discrete Fourier Transform matrix.
In other words a basis change from the rows of $x$ and $xz$ to the rows of $N$ and $H$
provides a more natural set of multidimensional axes in some contexts.
For $t$ a non-negative integer,
\beg N^{3t} \simeq I, \mf
N^{3t+1} \simeq N, \mf N^{3t+2} \simeq H, \mf N^{24} = I
\label{NPowers} \enspace,\eeg
so $N$ could be considered a 'generator' of $\{I,H,N\}$.
The $\{I,H,N\}^n$ transform set over $n$ binary variables
has been used to analyse the resistance of certain S-boxes to a form of generalised linear
approximation in
\cite{Par:SB}. It also defines the basis axes under which aperiodic
autocorrelation of boolean
functions is investigated in \cite{DanAPC}. The {\em{Negahadamard Transform}}, $\{N\}^n$, was introduced in
\cite{Par:Bent}. Constructions for boolean functions with
favourable spectral properties w.r.t. $\{H,N\}^n$
(amongst others) have been proposed in \cite{Par:LowPAR}, and \cite{Par:QE}
showed that boolean functions that are LU-equivalent to indicators for
distance-optimal binary error-correcting
codes yield favourable spectral properties w.r.t. $\{I,H\}^n$.

\subsection{A Spectral Derivation of LC}
\label{SpecDeriv}
We now re-derive LC by examining the repetitive action of $N$ on the vector form of
the graph states, interspersed with the actions of certain matrices from $D$.
We will show that, as with Lemma \ref{GEq}, these repeated actions not only generate the LC-orbit of
the graph, but also generate the $\{I,H,N\}^n$
transform spectra. The LC-orbit can be identified
with a subset of the flat transform spectra w.r.t. $\{I,H,N\}^n$.
Let $s = (-1)^{p({\bf{x}})}$, where $p({\bf{x}})$ is quadratic and represents a graph
$G$. Then the action of $N_v$ on $G$ is equivalent to $U_vs$, where:
$$ U_v \simeq U'_v = I \otimes \cdots \otimes I \otimes N \otimes I \otimes \cdots \otimes I \enspace,$$
where $N$ occurs at position $v$ in the tensor product decomposition.
Let us write $p({\bf{x}})$, uniquely, as,
$$ p({\bf{x}}) = x_v{\cal{N}}_v({\bf{x}}) + q({\bf{x}}) \enspace,$$
where $q({\bf{x}})$ and ${\cal{N}}_v({\bf{x}})$ are independent of $x_v$
(${\cal{N}}_v({\bf{x}})$ has nothing to do with the Negahadamard
kernel, $N_v$). We shall state a theorem that holds for $p({\bf{x}})$ of any degree,
not just quadratic, and then show that its specialisation to quadratic
$p({\bf{x}})$ gives the required single LC operation. Express
${\cal{N}}_v({\bf{x}})$ as the sum of $r$ monomials, $m_i({\bf{x}})$,
as follows,
$$ {\cal{N}}_v({\bf{x}}) = \sum_{i=0}^{r-1} m_i({\bf{x}}) \enspace.$$
For $p({\bf{x}})$ of any degree, the $m_i({\bf{x}})$ are of degree
$\le n - 1$. In the sequel we mix arithmetic, mod 2, and mod 4 so, to clarify
the formulas for equations that mix moduli, anything in square brackets is
computed $(\mo 2)$. The $\{0,1\}$ result is then embedded in $(\mo 4)$ arithmetic
for subsequent operations outside the square brackets.
We must also define,
$$ {\cal{N}}'_v({\bf{x}}) = \sum_{i=0}^{r-1} [m_i({\bf{x}})] \mf (\mo 4) \enspace.$$
\begin{thm}
Let $s' = U_vs$, where $s = (-1)^{p({\bf{x}})}$ and $s' = i^{p'({\bf{x}})}$.
Then,
\beg  p'({\bf{x}}) = 2\left[ p({\bf{x}}) +
\sum_{j\neq k}
m_j({\bf{x}})m_k({\bf{x}})  \right]+3{\cal{N}}'_v({\bf{x}})
 + 3[x_v]  \mf (\mo 4) \label{LC1} \enspace.\eeg
\label{thm2}
\end{thm}
\begin{proof}
Assign to $A$ and $B$ the evaluation of $p({\bf{x}})$ at $x_v = 0$ and
$x_v = 1$, respectively. Thus,
$$ A = p({\bf{x}})_{x_v = 0} = q({\bf{x}}) \enspace.$$
Similarly,
$$ B = p({\bf{x}})_{x_v = 1} = {\cal{N}}_v({\bf{x}}) + q({\bf{x}}) \enspace.$$
We need the following equality between mod 2 and mod 4 arithmetic.
\begin{lem}
$$\sum_{i=1}^n [A_i] \mz (\mo 4)\mf  = \mf [\sum_{i=1}^n A_i] + 2[\sum_{i\neq j} A_i A_j] \mz (\mo 4) \wh A_i \in {\mathbb{Z}}_2 \enspace.$$
\label{lem2} \end{lem}
Observe the following action of $N$:

\begin{small}
$$ \begin{array}{lcr}
\frac{1}{\sqrt{2}}\left ( \begin{array}{rr} 1 & i \\ 1 & -i \end{array} \right )
\left ( \begin{array}{r} 1 \\ 1 \end{array} \right ) & = &
w\left ( \begin{array}{r} 1 \\ -i \end{array} \right ) \\
\frac{1}{\sqrt{2}}\left ( \begin{array}{rr} 1 & i \\ 1 & -i \end{array} \right )
\left ( \begin{array}{r} -1 \\ 1 \end{array} \right ) & = &
w\left ( \begin{array}{r} i \\ -1 \end{array} \right ) \\
\frac{1}{\sqrt{2}}\left ( \begin{array}{rr} 1 & i \\ 1 & -i \end{array} \right )
\left ( \begin{array}{r} 1 \\ -1 \end{array} \right ) & = &
w\left ( \begin{array}{r} -i \\ 1 \end{array} \right ) \\
\frac{1}{\sqrt{2}}\left ( \begin{array}{rr} 1 & i \\ 1 & -i \end{array} \right )
\left ( \begin{array}{r} -1 \\ -1 \end{array} \right ) & = &
w\left ( \begin{array}{r} -1 \\ i \end{array} \right )
\end{array} $$
\end{small}
where $w = e^{2\pi i/8}$.
We ignore the global constant, $w$, so that $N$ maps $(-1)^{00}$ to $i^{03}$,
$(-1)^{10}$ to $i^{12}$, $(-1)^{01}$ to $i^{30}$ and $(-1)^{11}$ to $i^{21}$.
In general, for $A,B \in {\mathbb{Z}}_2$, $\al,\beta \in {\mathbb{Z}}_4$, $(-1)^{AB}$ is mapped by
$N_v$ to $i^{\al \beta}$, where,
$$ \begin{array}{lr}
\al = 2[AB] + [A] + 3[B] & (\mo 4)\\
\beta = 2[AB] + 3[A] + [B] + 3 & (\mo 4) 
\end{array} $$
Substituting the previous expressions for $A$ and $B$ into the above and making
use of Lemma \ref{lem2} gives,
$$ \begin{array}{lr}
\al({\bf{x}}) = 2[q({\bf{x}})] + 3[{\cal{N}}_v({\bf{x}})] & (\mo 4)\\
\beta({\bf{x}}) = 2[q({\bf{x}})] + [{\cal{N}}_v({\bf{x}})] + 3 & (\mo 4) 
\end{array} $$
$p'({\bf{x}})$ can now be written as,
$$ p'({\bf{x}}) = (3[x_v] + 1)\al({\bf{x}}) + [x_v]\beta({\bf{x}}) \mf (\mo 4) \enspace.$$
Substituting for $\al$ and $\beta$ gives,
$$ p'({\bf{x}}) = 2[q({\bf{x}})]+2[x_v{\cal{N}}_v({\bf{x}})]
+3[{\cal{N}}_v({\bf{x}})]
 + 3[x_v] \mf (\mo 4) $$
Applying Lemma \ref{lem2} to the term
$3[{\cal{N}}_v({\bf{x}})]$,
$$ 3[{\cal{N}}_v({\bf{x}})] = 2 \left[ \sum_{j\neq k}
m_j({\bf{x}})m_k({\bf{x}}) \right] + 3{\cal{N}}'_v({\bf{x}}) \mf (\mo 4) \enspace.$$
Furthermore, Lemma \ref{lem2} implies that,
$$  2\left[\sum_{i=1}^n A_i \right] (\mo 4) = 2 \sum_{i=1}^n [A_i] (\mo 4) \wh A_i \in {\mathbb{Z}}_2 \enspace.$$

\end{proof}

For $p({\bf{x}})$ a quadratic function, ${\cal{N}}_v({\bf{x}})$ has degree one,
so ${\cal{N}}'_v({\bf{x}})$ is a sum of degree-one terms over ${\mathbb{Z}}_4$. Therefore
the ${\mathbb{Z}}_4$ degree-one
terms, ${\cal{N}}'_v({\bf{x}})$ and $3[x_v]$, can be eliminated from
(\ref{LC1})
by appropriate subsequent action by members of $\{{\bf{D}}\}^n$ to $s'$. As all monomials,
$m_i({\bf{x}})$, are then of degree one,
(\ref{LC1}) reduces to,
\beg p'({\bf{x}}) \simeq p({\bf{x}}) +
\sum_{j,k \in {\cal{N}}_v, j \ne k} x_jx_k
 \mf (\mo 2) \label{LC2} \enspace.\eeg
(\ref{LC2}) precisely defines the action of a single LC operation at
vertex $v$ of $G$, where we have used $\simeq$ to mean that
$(-1)^{p'({\bf{x}})} = BU(-1)^{p({\bf{x}})}$, for some fully tensor-factorisable
matrix, $U$, and some $B \in \{{\bf{D}}\}^n$. 
As $p'({\bf{x}})$ is also quadratic boolean, we can
realise successive LC operations on chosen vertices in $G$ via successive actions of $N$
at these vertices, where each action of $N$
{\underline{must}} be interspersed with the action of a matrix from $\{{\bf{D}}\}^n$
to eliminate
${\mathbb{Z}}_4$-linear terms from (\ref{LC1}). In particular, one needs to
intersperse with tensor products of
$\begin{tiny} \left ( \begin{array}{ll} 1 & 0 \\ 0 & 1 \end{array} \right )
\end{tiny}$ and
$\begin{tiny} \left ( \begin{array}{ll} 1 & 0 \\ 0 & i \end{array} \right )
\end{tiny}$.
\begin{thm}
Given a graph, $G$, as represented by $s = (-1)^{p({\bf{x}})}$, with
$p({\bf{x}})$ quadratic,
the LC-orbit of $G$ comprises graphs which occur as a subset of the
spectra w.r.t. $\{I,H,N\}^n$
acting on $s$.
\label{LCinIHN}
\end{thm}
\begin{proof}
Define $D_1 \subset D$ such that
$$ D_1 =
\{\begin{tiny} \left ( \begin{array}{ll} a & 0 \\ 0 & b \end{array} \right )
\end{tiny},\begin{tiny} \left ( \begin{array}{ll} 0 & a \\ b & 0 \end{array} \right )
\end{tiny} \mz | \mz a = 1, b = \pm 1 \} \enspace.$$
Similarly, define $D_2 \subset D$ such that
$$ D_2 =
\{\begin{tiny} \left ( \begin{array}{ll} a & 0 \\ 0 & b \end{array} \right )
\end{tiny},\begin{tiny} \left ( \begin{array}{ll} 0 & a \\ b & 0 \end{array} \right )
\end{tiny} \mz | \mz a = 1, b = \pm i \}, \wh i^2 = -1 \enspace.$$
Then it is straightforward to establish that, for any
$\Delta_1,\Delta_1' \in D_1$, any $\Delta_2,\Delta_2' \in D_2$, and any
$c \in \{1,i,-1,-i\}$,
\beg \begin{array}{l}
N\Delta_1 = c \Delta_1'N \mf \mf  H\Delta_1 = c\Delta_1'H \\
N\Delta_2 = c \Delta_1H \mf \mf H\Delta_2 = c \Delta_1N \enspace.
\end{array}
\label{DMove} \eeg
Let $\Delta_* \in D_1 \bigcup D_2$. Then, for a vertex,
succesive applications of $\Delta_*N$ can, using (\ref{DMove}), be
re-expressed as,
$$ \prod (\Delta_*N) = c \Delta_* \prod N \simeq \prod N  \enspace.$$
But, from (\ref{NPowers}), successive powers of $N$ generate
$I$, $H$, or $N$, to within a final multiplication by a member of $D$.
It follows that successive LC actions on arbitrary vertices
can be described by the action on $s$ of a member of
the transform set, $\{I,H,N\}^n$, and therefore that the LC-orbit occurs
within the $\{I,H,N\}^n$ transform spectra of $s$.
\end{proof}

\subsection{LC on Hypergraphs}
For $p({\bf{x}})$ of degree $> 2$, ${\cal{N}}_v({\bf{x}})$ will
typically have degree higher than 1, and therefore the expansion of the sum will
contribute higher degree terms. For such a scenario
we can no longer eliminate the nonlinear and non-boolean term,
${\cal{N}}'_v({\bf{x}})$, from the right-hand side of (\ref{LC1}) by
subsequent actions from ${\bf{D}}$. Therefore, it is typically not possible to
iterate LC graphically beyond one step. We would like to
identify hypergraph equivalence w.r.t.
{\underline{local}} unitary transforms, in particular w.r.t. $\{I,H,N\}^n$.
Computations have shown that orbits
of boolean functions of degree $> 2$ and size greater than one do sometimes
exist with respect to $\{I,H,N\}^n$, although they appear to be significantly smaller in size
compared to orbits for the quadratic case \cite{DanAPC}.

{\em{An interesting open problem is to characterise a 'LC-like' equivalence for hypergraphs.}}

Further spectral symmetries of boolean functions w.r.t. $\{I,H,N\}^n$ are
discussed in Appendix B.

\section{Generalised Bent Properties of Boolean Functions}
%{I-Bent, Bent$_4$, I-Bent$_4$, LC-Bent, and ${\mathbb{Z}}_4$-Bent Boolean Functions}
\label{BentSection}
\subsection{Bent Boolean Functions}
A bent boolean function can be defined by using the WHT.
Let $p({\bf{x}})$ be our function over $n$ binary variables. Define
the WHT of $p({\bf{x}})$ by,
\beg P_{\bf{k}} =
2^{-n/2}\sum_{{\bf{x}} \in GF(2)^n} (-1)^{p({\bf{x}}) + {\bf{k \cdot x}}} \enspace, \label{WHT} \eeg
where ${\bf{x,k}} \in {\mbox{GF}}(2)^n$, and $\cdot$ implies the scalar product of vectors.

The WHT of $p({\bf{x}})$ can alternatively be defined as a
multiplication of the vector $(-1)^{p({\bf{x}})}$ by \linebreak 
$H \otimes H \otimes \ldots \otimes H$. Thus,
\beg P = 2^{-n/2}(H \otimes H \otimes \ldots \otimes H)(-1)^{p({\bf{x}})}
 = 2^{-n/2}(\bigotimes_{i=0}^{n-1} H)(-1)^{p({\bf{x}})} \enspace, \label{WHTMatrix} \eeg
where $P=(P_{(0,\ldots,0)},\ldots,P_{(1,\ldots,1)}) \in {\mathbb{C}}^{2^n}$. \newline
$p({\bf{x}})$ is defined to be {\em{bent}} if $|P_{\bf{k}}| = 1$
$\forall {\bf{k}}$,
in which case we say that $p({\bf{x}})$ has a {\em{flat spectra}} w.r.t.
the WHT. In other words, $p({\bf{x}})$ is bent if $P$ is {\em{flat}}.

Let $\Gamma$ be the binary adjacency
matrix associated to $p({\bf{x}})$ when $p({\bf{x}})$ is a quadratic.
\begin{lem}\cite{MacW:Cod}
$$ p({\bf{x}}) \m{ is {\em{bent}} } \mv \Leftrightarrow \mv \Gamma \m{ has maximum rank, mod 2} \enspace.$$
\label{MaxRank}
\end{lem}

It is well-known \cite{MacW:Cod} that all bent quadratics are equivalent under
affine transformation to the boolean function
$\left ( \sum_{i=0}^{\frac{n}{2}-1} x_{2i}x_{2i+1} \right ) + {\bf{c \cdot x}} + d$
for $n$ even, where ${\bf{c}} \in {\mbox{GF}}(2)^n$, and $d \in {\mbox{GF}}(2)$.
More generally, bent boolean functions only exist for
$n$ even. It is interesting to investigate other bent symmetries where
affine symmetry has been omitted. In particular, in the context of LC, we are interested
in the existence and number of flat spectra of boolean functions with respect to
the $\{H,N\}^n$-transform set ({\em{bent$_4$}}),
the $\{I,H\}^n$-transform set ({\em{I-bent}}), and
the $\{I,H,N\}^n$-transform set ({\em{I-bent$_4$}}).  
In the following subsections we investigate the bent$_4$, ${\mathbb{Z}}_4$-bent,
(Completely) I-bent, LC-bent, and (Completely) I-bent$_4$
properties of connected quadratic boolean functions,
where affine symmetry is omitted, and make some general statements about these properties
for more general boolean functions.

\subsection{Bent Properties with respect to $\{H,N\}^n$}
We now investigate certain spectral properties of boolean functions w.r.t.
$\{H,N\}^n$, where $\{H,N\}^n$ is the set of $2^n$ transforms
of the form $\bigotimes_{j \in {\bf{R_H}}} H_j
\bigotimes_{j \in {\bf{R_N}}} N_j,$ where the sets ${\bf{R_H}}$ and ${\bf{R_N}}$
partition $\{0,\ldots,n-1\}$.

The following is trivial to verify:
$$ p({\bf{x}}) \m{ is bent } \Leftrightarrow p({\bf{x}}) + {\bf{k \cdot x}} + d
 \mz \m{ is bent} \enspace,$$
where ${\bf{k}} \in {\mbox{GF}}(2)^n$ and $d \in {\mbox{GF}}(2)$.
In other words, if $p({\bf{x}})$ is bent then so
are all its affine offsets, mod 2.
However the above does not follow if one considers every possible ${\mathbb{Z}}_4$-linear
offset of the boolean function.
The WHT of $p({\bf{x}})$
with a ${\mathbb{Z}}_4$-linear offset can be defined as follows.
\beg P_{\bf{k},\bf{c}} = 2^{-n/2}\sum_{{\bf{x}} \in GF(2)^n}
(i)^{2[p({\bf{x}}) + {\bf{k \cdot x}}] + [{\bf{c \cdot x}}]}
\mf {\bf{k}},{\bf{c}} \in {\m{GF}}(2)^n \enspace. \label{WHT4} \eeg
\begin{df}
$$ p({\bf{x}}) \m{ is } \m{{\em{bent$_4$}} } \mv \Leftrightarrow \mv
\exists {\bf{c}} \m{ such that } |P_{{\bf{k},\bf{c}}}| = 1 \mf
\forall {\bf{k}} \in {\m{GF}}(2)^n \enspace.$$
\label{Bent4Def}
\end{df}

Let ${\bf{R_N}}$ and ${\bf{R_H}}$ partition $\{0,1,\ldots,n-1\}$.
Let,
$$ U = \bigotimes_{j \in {\bf{R_H}}} H_j \bigotimes_{j \in {\bf{R_N}}} N_j \enspace.$$
\beg s' = U(-1)^{p({\bf{x}})} \label{HNeq} \enspace.\eeg
\begin{lem}
$p({\bf{x}})$ is bent$_4$ if there exists one or more partitions,
${\bf{R_N}},{\bf{R_H}}$ such that $s'$ is {\em{flat}}.
\label{HN}
\end{lem}
\begin{proof}
The rows of $U$ can be described by $(i)^{f({\bf{x}})}$, where
${\bf{x}} = (x_0,x_1,\ldots,x_{n-1})$, where $f$ is linear,
$f : {\mbox{GF}}(2)^n \rightarrow {\mbox{GF}}(4)$, and the coefficient of $x_j$ in
$f \in \{0,2\}$ for
$j \in {\bf{R_H}}$ and $f \in \{1,3\}$ for $j \in {\bf{R_N}}$.
Therefore $s'$ can always, equivalently, be expressed
as $s' = (\bigotimes H)(i)^{2p[{\bf{x}}] + [f'({\bf{x}})]}$, where $f'$ is linear,
$f':{\mbox{GF}}(2)^n \rightarrow {\mbox{GF}}(2)$, and the coefficient of $x_j$ in $f'$ is $0$ for
$j \in {\bf{R_H}}$, and $1$ for $j \in {\bf{R_N}}$.
\end{proof}
An alternative way to define
the bent$_4$ property for $p({\bf{x}})$ quadratic
is via a modified form of the adjacency matrix.
\begin{lem}
For quadratic $p({\bf{x}})$,
$$ p({\bf{x}}) \m{ is bent$_4$ } \mv \Leftrightarrow \mv \Gamma_{\bf{v}}
\m{ has maximum rank, mod 2, for some } {\bf{v}} \in {\mbox{GF}}(2)^n \enspace.$$ \label{MaxRank4} \end{lem}
where $\Gamma_{\bf{v}}$ is a modified form of $\Gamma$ with $v_i$ in position
$[i,i]$, where ${\bf{v}} = (v_0,v_1,\ldots,v_{n-1})$.

\begin{proof}
We first show that the transform of $(-1)^{p({\bf{x}})}$ by tensor products
of $H$ and $N$ produces a flat spectra if and only if the associated periodic and
negaperiodic autocorrelation spectra have zero out-of-phase values.
We then show how these autocorrelation constraints lead directly to
constraints on the associated adjacency matrix.

Consider a function, $p$, of just one variable, $x_0$, and let
$s = (-1)^{p({x_0})}$. Define the periodic autocorrelation function
as follows,
$$ a_k = \sum_{x_0 \in {GF}(2)} (-1)^{p(x_0) + p(x_0 + k)}, \mf k \in {\mbox{GF}}(2) \enspace.$$
Then it is well-known that $s' = Hs$ is a flat spectrum if and only
if $a_k = 0$ for $k \ne 0$.

Define the negaperiodic autocorrelation function
as follows,
$$ b_k = \sum_{x_0 \in {GF}(2)} (-1)^{p(x_0) + p(x_0 + k) + k(x_0 + 1)}, \mf k \in {\mbox{GF}}(2) \enspace.$$
Then $s' = Ns$ is a flat spectrum if and only
if $b_k = 0$ for $k \ne 0$. (For $p$ a boolean function of just one variable,
$Hs$ is never flat and $Ns$ is always flat, but this
only holds for one variable).

We now elaborate on the above two claims.
Define $s(z) = s_0 + s_1z$, $a(z) = a_0 + a_1z$, and \linebreak $b(z) = b_0 + b_1z$.
Then the periodic and negaperiodic relationships between autocorrelation and
fourier spectra, as claimed above,
follow because periodic autocorrelation can be realised by
the polynomial multiplication, $a(z) = s(z)s(z^{-1})$ mod $(z^2 - 1)$,
with associated residue reduction, mod
$(z-1)$ and mod $(z+1)$, realised by
$s' = Hs =
\begin{tiny} \frac{1}{\sqrt{2}}\left ( \begin{array}{rr}
1 & 1 \\
1 & -1
\end{array} \right ) \end{tiny}s$ (with the Chinese Remainder Theorem
realised by $H^{\dag}s'$, where '$\dag$' means transpose conjugate).
By Parseval, $s'$ can only be flat if $a_1 = 0$.
Similarly, negaperiodic autocorrelation can be realised by the polynomial
multiplication, $b(z) = s(z)s(z^{-1})$ mod $(z^2 + 1)$,
with associated residue reduction, mod
$(z-i)$ and mod $(z+i)$, realised by
$s' = Ns =
\begin{tiny} \frac{1}{\sqrt{2}}\left ( \begin{array}{rr}
1 & i \\
1 & -i
\end{array} \right ) \end{tiny}s$ (with the Chinese Remainder Theorem
realised by $N^{\dag}s'$). By Parseval, $s'$ can only be flat if $b_1 = 0$.

We extend this autocorrelation $\leftrightarrow$ Fourier spectrum
duality
to $n$ binary variables by defining multivariate forms of the above polynomial
relationships. If we choose periodic autocorrelation for indices in ${\bf{R_H}}$ and negaperiodic
autocorrelation for indices in ${\bf{R_N}}$, we obtain the autocorrelation
spectra,
\beg A_{{\bf{k}},{\bf{R_H}},{\bf{R_N}}} =
\sum_{{\bf{x}} \in GF(2)^n} (-1)^{p({\bf{x}}) + p({\bf{x}}+ {\bf{k}})
+ \sum_{i=0}^{n-1} \chi_{_{\bf{R_N}}}(i)k_i (x_i + 1)} \enspace, \label{spectra} \eeg where
${\bf{k}}=(k_0,k_1,\ldots,k_{n-1}) \in {\mbox{GF}}(2)^n$, and $\chi_{_{\bf{R_N}}}(i)$
is the characteristic function of ${\bf{R_N}}$,
i.e, $$\chi_{_{\bf{R_N}}}(i)=\left\{\begin{array}{l}
1,\ i\in{\bf{R_N}}\\
0,\ i\notin{\bf{R_N}}\end{array}\right.$$
In polynomial terms, with ${\bf{z}} \in {\mbox{GF}}(2)^n$ and $s({\bf{z}}) =
\sum_{{\bf{j}} \in GF(2)^n} s_{\bf{j}} \prod_{i=0}^{n-1} z_i^{j_i}$, we have,
\beg \begin{array}{ll}
A_{{\bf{R_H}},{\bf{R_N}}}({\bf{z}}) & =
\displaystyle \sum_{{\bf{k}} \in GF(2)^n} A_{{\bf{k}},{\bf{R_H}},{\bf{R_N}}}
\displaystyle \prod_{i=0}^{n-1} z_i^{k_i} \\
 & = s(z_0,z_1,\ldots,z_{n-1})s(z_0^{-1},z_1^{-1},\ldots,z_{n-1}^{-1})
\mo \displaystyle \prod_{i=0}^{n-1} (z_i^2 - (-1)^{\chi_{\bf{R_N}}(i)}) \enspace.
\end{array} \label{ACFPol} \eeg

%Let ${\bf{k}} = (k_0,k_1,\ldots,k_{n-1})^T$ such that $k_i = 0$ if
%$i \in {\bf{R_H}}$ and $k_i = 1$ if $i \in {\bf{R_N}}$.
Then, by appealing to a multivariate version of Parseval's Theorem,
$s'$ as defined in (\ref{HNeq}) is flat if and only if
$A_{{\bf{k}},{\bf{R_H}},{\bf{R_N}}} = 0$,
$\forall\ {\bf{k}} \ne {\bf{0}}$.

These constraints on the autocorrelation coefficients of $s$ translate
to requiring a maximum rank property for a modified adjacency matrix,
as follows.
The condition $A_{\bf{k},{\bf{R_H}},{\bf{R_N}}} = 0$
for ${\bf{k}} \ne {\bf{0}}$ is equivalent to requiring that,
if we compare the function with its multidimensional periodic and negaperiodic
 rotations (but for the identity rotation), the remainder should be a balanced function.
When dealing with quadratic boolean functions, the remainder is always
linear or constant.
 This gives us a system of linear equations represented by the
 binary adjacency matrix, $\Gamma$, of $p({\bf{x}})$, with a modified
 diagonal, that is with $\Gamma_{i,i} = 1$ for all $i \in {\bf{R_N}}$, and
 $\Gamma_{i,i} = 0$ otherwise. Let
$$p(x_{0},x_{1},\ldots,x_{n-1})=a_{01}x_{0}x_{1}+a_{02}x_{0}x_{2}+\cdots+a_{ij}x_{i}x_{j}+\cdots+a_{n-2,n-1}x_{n-2}x_{n-1} \enspace.$$ 

%This, expressed as the adjacency matrix of the graph, is:

%\begin{small}
%$$\Gamma=\left(\begin{array}{ccccc}
%0 & a_{01} & a_{02} & \ldots & a_{0,n-1}\\
%a_{01} & 0 & a_{12} & \ldots & a_{1,n-1}\\
%a_{02} & a_{12} & 0 & \ldots & a_{2,n-1}\\
%\vdots &\vdots  &\vdots  & \ddots & \vdots\\
%a_{0,n-1} & a_{1,n-1} & a_{2,n-1} & \ldots & 0\end{array}\right) \enspace.$$
%\end{small}

%Substituting in the ANF, we see that
%$$\begin{array}{lcl}
%p(x_0+k_0,\ldots,x_{n-1}+k_{n-1}) & = &
%p(x_0,\ldots,x_{n-1})+
%k_0(a_{01}x_{1}+a_{02}x_{2}+\cdots+a_{0,n-1}x_{n-1})\\
%& + &  k_1(a_{01}x_{0}+a_{02}x_{2}+\cdots+a_{0,n-1}x_{n-1})+\cdots\\
%& + & k_{n-1}(a_{0,n-1}x_{0}+a_{1,n-1}x_{2}+\cdots+a_{n-2,n-1}x_{n-2}) \enspace.
%\end{array} $$
Therefore, 
$$\begin{array}{lcl}
p({\bf{x}}) + p({\bf{x}} + {\bf{k}})+\sum_{i=0}^{n-1} \chi_{_{\bf{R_N}}}(i)k_i x_i & = &  k_0(\chi_{_{\bf{R_N}}}(0)x_0+a_{01}x_{1}+a_{02}x_{2}+\cdots+a_{0,n-1}x_{n-1})\\
& + & k_1(a_{01}x_{0}+\chi_{_{\bf{R_N}}}(1)x_1+a_{02}x_{2}+\cdots+a_{0,n-1}x_{n-1})+\cdots\\
& + & k_{n-1}(a_{0,n-1}x_{0}+\cdots+a_{n-2,n-1}x_{n-2}+\chi_{_{\bf{R_N}}}(n-1)x_{n-1}) \enspace.
\end{array}$$ This is equal to:
$$\begin{array}{l}
x_{0}(\chi_{_{\bf{R_N}}}(0)k_0+a_{01}k_1+\cdots+a_{0n}k_n)+x_{1}(a_{01}k_0+\chi_{_{\bf{R_N}}}(1)k_1+
\cdots a_{1,n-1}k_{n-1})\\
+ \cdots+x_{n-1}(a_{0,n-1}k_0+a_{1,n-1}k_1+\cdots+a_{n-2,n-1}k_{n-2}+\chi_{_{\bf{R_N}}}(n-1)k_{n-1}) \enspace,
\end{array}$$
which is balanced unless constant.
The constant $\sum_{i=0}^{n-1} \chi_{_{\bf{R_N}}}(i)k_i$ will not play
any role in the equation $A_{\bf{k}}=0$, and can be ignored.
We have the the following system of equations:
$$\begin{array}{cl}
&\chi_{_{\bf{R_N}}}(0)k_0+a_{01}k_1+a_{02}k_2+\cdots+a_{0,n-1}k_{n-1}=0\\
&a_{01}k_0+\chi_{_{\bf{R_N}}}(1)k_1+a_{12}k_2+\cdots+a_{1,n-1}k_{n-1}=0\\
&.................................................................................\\
&a_{0,n-1}k_0+a_{1,n-1}k_1+\cdots+a_{n-2,n-1}k_{n-2}+\chi_{_{\bf{R_N}}}(n-1)k_{n-1}=0 \enspace.
\end{array}$$
Writing this system as a matrix, we have:

\begin{small}
$$\left(\begin{array}{ccccc}
\chi_{_{\bf{R_N}}}(0) & a_{01} & a_{02} & \ldots & a_{0,n-1}\\
a_{01} & \chi_{_{\bf{R_N}}}(1) & a_{12} & \ldots & a_{1,n-1}\\
a_{02} & a_{12} & \chi_{_{\bf{R_N}}}(2) & \ldots & a_{2,n-1}\\
\vdots &\vdots  &\vdots  & \ddots & \vdots\\
a_{0,n-1} & a_{1,n-1} & a_{2,n-1} & \ldots & \chi_{_{\bf{R_N}}}(n-1)\end{array}\right) \enspace.$$
\end{small}

This is a modification of $\Gamma$, with 1 or 0 in position $i$ of the
diagonal depending on whether $i\in{\bf{R_N}}$ or $i\in{\bf{R_H}}$.
\end{proof}
%{\bf Obs.}: That implies that the adjacency matrix can never be full rank if $n$ is odd.

In general,
$$ p({\bf{x}})  \m{ is bent }
\begin{array}{l} \Rightarrow \\ 
{\not \Leftarrow} \end{array} \mf p({\bf{x}}) \m{ is bent$_4$} \enspace.$$

\begin{thm}
All boolean functions of degree $\le 2$ are bent$_4$.
\label{Z4Bent}
\end{thm}
\begin{proof}
Degree zero and degree one functions are trivial. Consider the adjacency matrix, $\Gamma$, associated with the quadratic boolean
function, $p({\bf{x}})$. We now prove that $\Gamma_{\bf{v}}$ has maximum rank (mod 2)
for at least one choice of ${\bf{v}}$, where $\Gamma_{\bf{v}} = \Gamma + \m{diag}({\bf{v}})$ as
before.
Let $M$ be the minor associated with the first entry of $\Gamma$; in other words, let
\begin{small}
${\Gamma}= \begin{tiny} \left ( \begin{array}{ll}
0 & \ \\
\  & M
\end{array} \right ) \end{tiny}  \enspace.$
\end{small}

We prove by induction that there exists at
least one choice of $v$ such that $\Gamma_{\bf{v}}$ has maximum rank (mod 2).
The theorem is true for $n=2$: in this case,
\begin{small} 
${\Gamma}= \begin{tiny} \left ( \begin{array}{ll}
0 & a \\
a & 0
\end{array} \right ) \end{tiny} \enspace.$
\end{small}

Then, either det$({\Gamma})=1$, in which case we choose ${\bf{v}}=(0,0)$, or we
have $a=0$ (empty graph).
In the last case we choose ${\bf{v}}=(1,1)$, so $det(\Gamma_{\bf{v}})=1+a=1$.
Suppose the theorem is true for $n-1$ variables. We will see that it is true for $n$ variables.
If the determinant of ${\Gamma}$ is 1 we take ${\bf{v}}=(0,\ldots,0)$ and we are done. 
If det$({\Gamma})=0$, then we have two cases:
\begin{itemize}
\item det$(M)=1$: Take ${\bf{v}}=(1,0,\ldots,0)$.
\item $det(M)=0$: By the induction hypothesis
there is at least one choice of ${\bf{v}}(M) \in {\mbox{GF}}(2)^{n-1}$, where
${\bf{v}}(M)=(v_1,\ldots,v_{n-1})$ such that $M_{{\bf{v}}(M)}$ has full rank.
Let ${\bf{v}}'=(0,v_1,\ldots,v_{n-1}) \in {\mbox{GF}}(2)^n$. If
det$({\Gamma_{\bf{v'}}}) = 1$ we have finished.
If det$({\Gamma_{\bf{v'}}}) = 0$ we are in the first case again, so we take ${\bf{v}}=(1,v_1,\ldots,v_{n-1})$, and we are done.
\end{itemize}
The theorem follows from lemma \ref{MaxRank4}.\end{proof}
{\bf{Remark:}} Theorem \ref{Z4Bent} is true even for boolean functions associated
with non-connected or empty graphs.

\begin{lem}
Not all boolean functions of degree $ > 2$ are bent$_4$.
\label{NotAllBent4}
\end{lem}
\begin{proof}
Counter-example -
by computation there are no bent$_4$ cubics of three variables.
\end{proof}
Further computations show that there are
no bent$_4$ boolean functions of four variables of degree $> 2$.
Similarly, there are only $252336$ bent$_4$ cubic boolean
functions in five variables (out of a possible $2^{20} - 2^{10}$, not
including affine offsets),
and no bent$_4$ boolean functions of degree $\ge 4$ in five variables.
bent$_4$ cubics of six variables do exist.
Lemma \ref{NotAllBent4} identifies an open problem:
\cb {\em{ What is the maximum algebraic degree of a bent$_4$ boolean function
of $n$ variables?}}
\ce

\begin{df}
$$ p({\bf{x}}) \m{ is } {\em{{\mathbb{Z}}_4\m{-bent }}} \Leftrightarrow
|P_{{\bf{k}},\bf{c}}| = 1 \mf \forall {\bf{c,k}} \in {\mbox{GF}}(2)^n \enspace.$$
\label{Z4BentDef}
\end{df}
The definition requires that {\bf{all}} ${\mathbb{Z}}_4$-linear offsets of the boolean function,
$p({\bf{x}})$, are flat w.r.t. the WHT. WE prove that no such boolean
functions exist, first for all boolean
functions of degree $\le 2$, and then for all boolean functions.
\begin{thm}
There are no ${\mathbb{Z}}_4$-bent boolean functions of degree $\le 2$.
\label{NoZ4BentQuad}
\end{thm}
\begin{proof}
This is trivial for degree zero and degree one functions.
Consider the adjacency matrix, $\Gamma$, associated with the quadratic boolean
function, $p({\bf{x}})$. The theorem is equivalent to proving that there is a ${\bf{v}}$ such that
$\Gamma_{\bf{v}}$ has rank  less than maximal. Then:
\begin{enumerate}
\item if $p({\bf{x}})$ is not bent, then we take ${\bf{v}}=(0,\ldots,0)$ and we are done.
\item if $p({\bf{x}})$ is bent, we take $M$ as in the proof for Theorem \ref{Z4Bent}.
If $det(M)=1$, we take 
${\bf{v}}=(1,0,\ldots,0)$ and we are done; if $det(M)=0$, modify the diagonal as in the proof
for Theorem \ref{Z4Bent}. If the determinant of the new matrix is equal to $0$, we are done; if not,
we are in case 1.
\end{enumerate}

\end{proof}

\begin{thm}
There are no ${\mathbb{Z}}_4$-bent boolean functions.
\label{NoZ4Bent}
\end{thm}
\begin{proof}
Consider the proof of Lemma \ref{MaxRank4}. We have established that, for a fixed choice
of ${\bf{R_H}}$ and ${\bf{R_N}}$,
$s'$, as defined in (\ref{HNeq}), is flat if and
only if $A_{{\bf{k}},{\bf{R_H}},{\bf{R_N}}} = 0$, $\forall {\bf{k}}$,
${\bf{k \ne 0}}$. Therefore $p({\bf{x}})$ is ${\mathbb{Z}}_4$-bent iff
$A_{{\bf{k}},{\bf{R_H}},{\bf{R_N}}} = 0$, $\forall {\bf{k}}$,
${\bf{k \ne 0}}$, for {\bf{all}} partitions $\{{\bf{R_H,R_N}}\}$. In particular,
if $p({\bf{x}})$ is ${\mathbb{Z}}_4$-bent,
then the polynomials, $A_{{\bf{R_H}},{\bf{R_N}}}({\bf{z}})$,
as defined in
(\ref{ACFPol}), satisfy $A_{{\bf{R_H}},{\bf{R_N}}}({\bf{z}}) = 2^n$ for all
choices of ${\bf{R_H}}$ and ${\bf{R_N}}$
(i.e. their out-of-phase coefficients are all zero). By the
Chinese Remainder Theorem (CRT) we can combine these polynomials for
each choice of ${\bf{R_H}}$ and ${\bf{R_N}}$ to construct the polynomial,
\beg r({\bf{z}}) \mo \prod_{j=0}^n (z_j^4 - 1) =
\m{CRT}\{A_{{\bf{R_H}},{\bf{R_N}}}({\bf{z}}) \mt | \mt \forall {\bf{R_H}},{\bf{R_N}}\}
\label{FullCRT} \enspace,\eeg
where $r({\bf{z}}) =
s(z_0,z_1,\ldots,z_{n-1})s(z_0^{-1},z_1^{-1},\ldots,z_{n-1}^{-1})$.

But as $r({\bf{z}})$ comprises monomials containing only $z_i^{-1},z_i^0,z_i^1$, the modular
restriction in (\ref{FullCRT}) has no effect on coefficient magnitudes, and
$$  r({\bf{z}}) \equiv r({\bf{z}}) \mo \prod_{j=0}^n (z_j^4 - 1) \enspace.$$
to within a multiplication of the coefficients by $\pm 1$.
It follows, by application of the CRT to (\ref{FullCRT}) that, if
$A_{{\bf{R_H}},{\bf{R_N}}}({\bf{z}}) = 2^n$,
$\forall {\bf{R_H}},{\bf{R_N}}$, then $r({\bf{z}}) = 2^n$ also, i.e.
$r({\bf{z}})$ is integer.
But this is impossible as the coefficients of the maximum degree terms,
$\prod_j z_j^{-1^{u_j}}$, $u_j \in {\mathbb{Z}}_2$, in
$r({\bf{z}})$ can never be zero, but are always $\pm 1$.
Therefore $p({\bf{x}})$ can never be ${\mathbb{Z}}_4$-bent.
\end{proof}

{\bf{Remark: }} Although we proved for boolean functions, it is
possible to generalise the proof so as to state that no function from
${\mbox{GF}}(2)^n \rightarrow {\mbox{GF}}(q)$ can be ${\mathbb{Z}}_4$-bent,
for any even integer $q$.

\subsection{Bent Properties with respect to $\{I,H\}^n$}
We now investigate certain spectral properties of boolean functions w.r.t.
$\{I,H\}^n$, where $\{I,H\}^n$ is the set of $2^n$ transforms
of the form $\bigotimes_{j \in {\bf{R_I}}} I_j \bigotimes_{j \in {\bf{R_H}}} H_j,$
where the sets ${\bf{R_I}}$ and ${\bf{R_H}}$  partition
$\{0,\ldots,n-1\}$. \cite{Par:QE} has investigated other spectral
properties w.r.t. $\{I,H\}^n$, such as {\em{weight hierarchy}} if the graph is bipartite.

The WHT of the subspace of a function from ${\mbox{GF}}(2)^n$ to ${\mbox{GF}}(2)$,
obtained by fixing a subset, ${\bf{R_I}}$, of the input variables,
can be defined as follows. Let ${\bf{\theta }} \in {\mbox{GF}}(2)^n$ be such that
$\theta_j = 1$ iff $j \in {\bf{R_I}}$. Let ${\bf{r \preceq \theta }}$, where '$\preceq$'
means that $\theta$ '{\em{covers}}' ${\bf{r}}$, i.e. $r_i \le \theta_i$, $\forall i$.
Then,
\beg P_{\bf{k},\bf{r},\bf{\theta}} =
2^{-(n - \m{wt}(\theta))/2}\sum_{{\bf{x}} = {\bf{r + y}} | {\bf{y}} \preceq {\bar{\theta}}}
(-1)^{p({\bf{x}}) + {\bf{k \cdot x}}}
\mf {\bf{k}} \preceq {\bar{\theta}},  {\bf{r}} \preceq \bf{\theta} \label{IHEQU} \enspace.\eeg
\begin{df}
$$ p({\bf{x}}) \m{ is } \m{{\em{I-bent}} } \mv \Leftrightarrow \mv
\exists \theta \m{ such that } |P_{{\bf{k},\bf{r},\bf{\theta}}}| = 1 \mf
\forall {\bf{k}} \preceq {\bf{\bar{\theta}}}, \forall \bf{r} \preceq \bf{\theta} \enspace,$$
where $\m{wt}(\bf{\theta}) < n$.
\label{IBentDef}
\end{df}

Let
\beg U = \bigotimes_{j \in {\bf{R_I}}} I_j \bigotimes_{j \in {\bf{R_H}}} H_j \enspace.
\label{UIH} \eeg
\beg s' = U(-1)^{p({\bf{x}})} \label{IHeq} \enspace.\eeg
\begin{df}
$p({\bf{x}})$ is I-bent if there exist one or more partitions,
${\bf{R_I}},{\bf{R_H}}$ such that $s'$ is flat, where $|{\bf{R_I}}| < n$.
\label{DefIH}
\end{df}

An alternative way to define
the I-bent property of $p({\bf{x}})$
is via its associated adjacency matrix, $\Gamma$.
Let $\Gamma_I$ be the adjacency matrix obtained from $\Gamma$ by deleting all rows
and columns of $\Gamma$ with indices in ${\bf{R_I}}$. 
\begin{lem}
For quadratic $p({\bf{x}})$,
$$ p({\bf{x}}) \m{ is I-bent } \mv \Leftrightarrow \mv \Gamma_I
\m{ has maximum rank, mod 2} $$
for one or more choices of ${\bf{R_I}}$ where $|{\bf{R_I}}| < n$.
\label{MaxRankIH}
\end{lem}

In general,
$$ p({\bf{x}})  \m{ is bent }
\begin{array}{l} \Rightarrow \\ 
{\not \Leftarrow} \end{array} \mf p({\bf{x}}) \m{ is I-bent} \enspace.$$

\begin{thm}
All boolean functions in two or more variables of degree $\le 2$ are I-bent.
\label{I-Bent}
\end{thm}
\begin{proof}
Degree zero and degree one functions are trivial.
It is easy to show that all
quadratic boolean functions of 2 variables are I-bent. The theorem follows
by observing that all adjacency matrices, $\Gamma$,
representing quadratic functions of $n > 2$ variables contain $2 \times 2$
submatrices, obtained from $\Gamma$ by deleting all rows and columns of $\Gamma$
with indices ${\bf{R_I}}$, for $|{\bf{R_I}}| = n - 2$.
\end{proof}

\begin{lem}
Not all boolean functions of degree $ > 2$ are I-bent.
\label{NotAllIBent}
\end{lem}
\begin{proof}
Counter-example - by computation there are no I-bent cubics of three variables.
\end{proof}
Further computations show that there are only 416
I-bent cubics in four variables, and no
I-bent quartics in four variables. There are only 442640 I-bent cubics,
only 1756160 I-bent quartics in five variables, and no I-bent quintics in
five variables. I-bent cubics in six variables do exist.
Lemma \ref{NotAllIBent} indicates an open problem:
\cb
{\em{What is the maximum algebraic degree of an I-bent boolean function of $n$ variables?}}
\ce

\begin{df}
$$ p({\bf{x}}) \m{ is } \m{{\em{Completely I-bent}} } \Leftrightarrow
|P_{{\bf{k}},\bf{r},\bf{\theta}}| = 1 \mf \forall \bf{\theta},{\bf{k}},{\bf{r}}, \mv
{\bf{k}} \preceq {\bar{\theta}}, {\bf{r}} \preceq \theta \enspace.$$
\label{CIBentDef}
\end{df}
\begin{thm}
There are no Completely I-bent boolean functions.
\label{NoCIBent}
\end{thm}
\begin{proof}
Let $s = (-1)^{p({\bf{x}})}$. Let $|{\bf{R_I}}| = n - 1$. Then for $U$ as defined in
(\ref{UIH}), $s'$ cannot be flat.
\end{proof}

\subsection{Bent Properties with respect to $\{I,H,N\}^n$}
The $\{H,N\}^{n - |{\bf{R_I}}|}$ set of transforms
of the subspace of a function from ${\mbox{GF}}(2)^n$ to ${\mbox{GF}}(2)$,
obtained by fixing a subset, ${\bf{R_I}}$, of the input variables,
is defined as follows. Let ${\bf{\theta }} \in {\mbox{GF}}(2)^n$ be such that
$\theta_j = 1$ iff $j \in {\bf{R_I}}$. Let ${\bf{r \preceq \theta }}$.
Then,
\beg P_{\bf{k},\bf{c},\bf{r},{\bf{\theta }}} =
2^{-(n - \m{wt}(\theta))/2}\sum_{{\bf{x}} = {\bf{r + y}} | {\bf{y}} \preceq {\bar{\theta}}}
(i)^{2[p({\bf{x}}) + {\bf{k \cdot x}}] + [{\bf{c \cdot x}}]}
\mf {\bf{k,c}} \preceq {\bar{\theta}},  {\bf{r}} \preceq \theta \enspace. \label{IHNEQU} \eeg
\begin{df}
$$ p({\bf{x}}) \m{ is } \m{{\em{I-bent$_4$}} } \mv \Leftrightarrow \mv
\exists {\bf{c}}, {\bf{\theta }} \m{ such that }
|P_{{\bf{k},\bf{c},\bf{r},{\bf{\theta }}}}| = 1 \mf
\forall {\bf{k}} \preceq {\bar{\bf{\theta }}}, \forall \bf{r} \preceq {\bf{\theta }} \enspace,$$
where $\m{wt}({\bf{\theta }}) < n$.
\label{I-Bent4Def}
\end{df}

Let ${\bf{R_I}}$, ${\bf{R_H}}$ and ${\bf{R_N}}$ partition
$\{0,1,\ldots,n-1\}$.
Let,
\beg U = \bigotimes_{j \in {\bf{R_I}}} I_j \bigotimes_{j \in {\bf{R_H}}} H_j
\bigotimes_{j \in {\bf{R_N}}} N_j \enspace.
\label{UIHN} \eeg
\beg s' = U(-1)^{p({\bf{x}})} \enspace. \label{IHNeq} \eeg
\begin{lem}
$p({\bf{x}})$ is I-bent$_4$ if there exists one or more partitions,
${\bf{R_I}},{\bf{R_H}},{\bf{R_N}}$ such that $s'$ is {\em{flat}},
where $|{\bf{R_I}}| < n$.
\label{IHN}
\end{lem}
As a generalization of
(\ref{spectra}), we get flat spectra for one or more partitions ${\bf{R_I}},{\bf{R_H}},{\bf{R_N}}$
iff
$$A_{k,{\bf{R_I}},{\bf{R_H}},{\bf{R_N}}}=
\sum_{{\bf x} = {\bf{r + y}} | {\bf{y}} \preceq {\bar{\bf{\theta }}}}(-1)^{p({\bf{x}}) +
p({\bf{x}}+ {\bf{k}})
+ \sum_{i=0}^{n-1} \chi_{_{\bf{R_N}}}(i)k_i (x_i + 1)}=0, \mf \forall {\bf{k}}\neq{\bf{0}} \enspace,$$
where $\theta_j = 1$ iff $j \in {\bf{R_I}}$, ${\bf{r}} \preceq \theta$, and
$r_j = k_j$ if $j \in {\bf{R_I}}$.

An alternative way to define
the I-bent$_4$ property when $p({\bf{x}})$ is
quadratic is via its associated adjacency matrix, $\Gamma$.
Let $\Gamma_{I,{\bf{v}}}$ be the matrix obtained from $\Gamma_{\bf{v}}$ when
we erase the $i^{th}$ row and column if $i\in{\bf{R_I}}$.
\begin{lem}
For quadratic $p({\bf{x}})$,
$$ p({\bf{x}}) \m{ is I-bent$_4$ } \mv \Leftrightarrow \mv \Gamma_{I,{\bf{v}}}
\m{ has maximum rank, mod 2}, \wh {\bf{v}} \preceq \bf{\bar{\theta}} $$
for one or more choices of ${\bf{v}}$ and $\theta$ where $\m{wt}(\theta) < n$.
\label{MaxRankIHN}
\end{lem}

In general,
$$
p({\bf{x}})  \m{ is bent }
\begin{array}{l} \Rightarrow \\ 
{\not \Leftarrow} \end{array}
\begin{array}{l}
\mf p({\bf{x}}) \m{ is bent$_4$} \\
\mf p({\bf{x}}) \m{ is I-bent}
\end{array}
\begin{array}{l} \Rightarrow \\ 
{\not \Leftarrow} \end{array}
\mf p({\bf{x}}) \m{ is I-bent$_4$} \enspace.
$$

\begin{thm}
All boolean functions of degree $\le 2$ are I-bent$_4$.
\label{I-Bent4}
\end{thm}
\begin{proof}
Follows from Theorems \ref{Z4Bent} and \ref{I-Bent}.
\end{proof}

\begin{lem}
All boolean functions are I-bent$_4$.
\label{NotAllIBent4}
\end{lem}
\begin{proof}
From Theorem \ref{thm2}, the action of a single
$U_v$ on a boolean function, $p({\bf{x}})$, of any degree,
always gives a flat output spectra, for any value of $v$.
This gives (at least) $n$ flat spectra for any boolean function.
\end{proof}

\begin{df}
$$ p({\bf{x}}) \m{ is } \m{{\em{Completely I-bent$_4$}} } \Leftrightarrow
|P_{{\bf{k}},{\bf{c}},\bf{r},\bf{\theta}}| = 1 \mf \forall
\bf{\theta},{\bf{c}},{\bf{k}},{\bf{r}}, \mv
{\bf{k}},{\bf{c}} \preceq {\bar{\theta}}, {\bf{r}} \preceq \theta \enspace.$$
\label{CIBent4Def}
\end{df}
\begin{thm}
There are no completely I-bent$_4$ boolean functions.
\label{NoCIBent4}
\end{thm}
\begin{proof}
Follows from theorems \ref{NoZ4Bent} or \ref{NoCIBent}.
\end{proof}

It is natural to ask whether, for a
given quadratic,
$p({\bf{x}})$, there exists at least one member of its LC-orbit
which is bent. If so, then
we state that the graph state, $p({\bf{x}})$, and its associated LC-orbit,
is {\em{LC-bent}}. More formally,
\begin{df}
The graph state, $p({\bf{x}})$ (a quadratic boolean function), and its
associated LC-orbit is {\em{LC-bent}}
if $\exists\ p'({\bf{x}})$ such that
$p'({\bf{x}}) \in \m{LC-orbit}(p({\bf{x}}))$, and such
that $p'({\bf{x}})$ is bent.
\label{LC-Bent}
\end{df}
For example, the bent function $x_0x_1 + x_0x_2 + x_0x_3 + x_1x_2 + x_1x_3 + x_2x_3$ is in the
same LC-orbit as $x_0x_1 + x_0x_2 + x_0x_3$ so, although
$x_0x_1 + x_0x_2 + x_0x_3$ is not bent,
it is LC-bent.

In general, for $p({\bf{x}})$ quadratic,
$$ p({\bf{x}})  \m{ is bent}
\begin{array}{l} \Rightarrow \\ 
{\not \Leftarrow} \end{array} \mf p({\bf{x}}) \m{ is LC-bent} \enspace.$$

\begin{thm}
Not all quadratic boolean functions are LC-bent.
\label{notallLCBent}
\end{thm}
\begin{proof}
By computation, the LC-orbit associated with the $n = 6$-variable
boolean function,
$x_0x_4 + x_1x_5 + x_2x_5 + x_3x_4 + x_4x_5$ is not LC-bent.
\end{proof}
By computation it was found that all quadratic boolean functions of $n \le 5$
variables are LC-bent.
Table I lists orbit representatives for those orbits which are not LC-bent, for
$n = 2$ to $9$, and provides a summary for $n = 10$,
where the boolean functions are abbreviated so
that, say, $ab,de,fg$ is short for $x_ax_b + x_dx_e + x_fx_g$.
For those orbits which are
not LC-bent we provide the maximum rank satisfied by a graph within the orbit.

\begin{table}[htb]
\cb
\begin{small}
\begin{tabular}{|c|c|c|} \hline
$n$   & ANF for the orbit representative & Max. Rank within Orbit \\ \hline \hline
2-5   &  -                               &  - \\
6     & 04,15,25,34,45                   &  4 \\
7     &  -                               &  -  \\
8     & 07,17,27,37,46,56,67             &  6  \\
      & 06,17,27,37,46,56,67             &  6  \\
      & 07,17,25,36,46,57,67             &  6  \\
      & 06,17,27,36,45,46,47,56,57,67    &  6  \\
      & 07,16,26,35,45,47,67             &  6  \\
9     & 08,18,28,38,47,57,67,78          &  6  \\
      & 08,18,26,37,47,56,68,78          &  6  \\
10    & 08,19,29,39,49,58,68,78,89       &  6  \\
      &     51 other orbits              &  8  \\ \hline    
\end{tabular}
\caption{Representatives for all LC-Orbits which are not LC-bent for $n = 2$ to $10$}
\end{small}
\ce
\label{notLCBent}
\end{table}

\section{Conclusion}
This paper has examined the spectral properties of boolean functions with
respect to the transform set formed by tensor products of
the identity, $I$, the Walsh-Hadamard kernel, $H$, and the Negahadamard
kernel, $N$ (the $\{I,H,N\}^n$ transform set).
In particular, the idea of a bent boolean function was generalised
in a number of ways to $\{I,H,N\}^n$.
Various theorems about the
generalised bent properties of boolean functions were established.
It was shown how a quadratic
boolean function maps to a graph and it was shown how the local
unitary equivalence of these graphs can be realised by successive
application of the LC operation - Local Complementation - or,
alternatively, by identifying a
subset of the flat spectra with respect to $\{I,H,N\}^n$.
For quadratic boolean functions it was further shown how the $\{I,H,N\}^n$ set
of transform spectra could be
characterised by looking at the ranks of suitably modified versions of the
adjacency matrix. In the second part of the paper, we will apply this method to
enumerate the flat spectra w.r.t. $\{I,H\}^n$, $\{H,N\}^n$ and $\{I,H,N\}^n$ for
certain concrete functions \cite{RP:BCII}.

\newpage

\section{Appendix A - Various Interpretations of the Graph States}
\label{graphstates}
In this section we briefly characterise graph states.

\subsection{Interpretation as a Quantum Error Correcting Code}
Let $E$ be a $2n$-dimensional binary vector space, whose elements are written
as $(a|b)$, where \linebreak $a,\ b\ \in \mbox{GF}(2)^n$, and $E$ is equiped with the (symplectic) inner product
$((a|b),(a'|b'))=a\cdot b'+a'\cdot b$.
Define the {\em weight} of $(a|b)=(a_1,\ldots,a_n|b_1,\ldots,b_n)$ as the number
of coordinates $i$ such that at least one of the $a_i$ or $b_i$ is 1.
The distance between two elements $(a|b)$ and $(a'|b')$ is defined to be the
weight of their difference.

\begin{thm} \label{bin} \cite{Cald:Qua}
Let $S$ be a $(n-k)$ - dimensional linear subspace of
$E$, contained in its dual $S^\perp$ (with respect to the inner product),
such that there are no vectors of weight $< d$ in $S\setminus S^\perp$.
By taking an eigenspace of $S$ (for any chosen linear character) we
obtain a quantum error-correcting code mapping $k$ qubits to $n$ qubits that
corrects $[(d-1)/2]$ errors. Such a code is called an {\em additive
quantum error-correcting code (QECC)}, and is described by its
parameters, $[[n,k,d]]$, where $d$ is the {\em minimal distance} of
the code.
\end{thm}
We show, later, that a $[[n,0,d]]$ QECC can be represented by a graph.
First we re-express the QECC as a $\GF(4)$ additive code.

\subsection{Interpretation as a $\GF(4)$ Additive Code}
From \cite{Cald:Qua} we see how to interpret the binary space $E$ as the
space GF(4)$^n$ and thereby how to derive a QECC from an additive (classical)
code over GF(4)$^n$. Let $GF(4)=\{0,1,\omega,\bar{\omega}\}$, with $\omega^2=\omega+1$, $\omega^3=1$;
and conjugation defined by $\bar{\omega}=\omega^2=\omega+1$. The {\em Hamming
weight} of a vector in GF(4)$^n$, written $wt(u)$, is the number of non-zero
components, and the {\em Hamming distance} between $u,u'\in$ GF(4)$^n$ is
dist$(u,u')=wt(u + u')$. Define the {\em trace function} as:
$tr(x): GF(4) \rightarrow GF(2),\ tr(x)=x+\bar{x}$. To each vector
$v=(a|b)\in E$ we associate the vector $\phi(v)=a\omega+b\bar{\omega}$.
The weight of $v$ is the Hamming weight
of $\phi(v)$, and the distance between two vectors in $E$ is the Hamming
distance of their images. If $S$ is a subspace of $E$ then $C=\phi(S)$
is a subset of GF$(4)^n$ that is closed under addition (defining thus an
additive code). The {\em trace inner product} of
$u,v\in \mbox{ GF}(4)^n$ is
$$u\star v=Tr(u\cdot \bar{v})=\sum_{i=1}^n(u_i\bar{v_i}+\bar{u_i}v_i) \enspace,$$
Define the {\em dual code} $C^\perp$ as
$$C^\perp=\{u\in \mbox{ GF}(4)^n:u\star v=0\ \forall v\in C\} \enspace.$$
Now one can reformulate Theorem \ref{bin}.
\begin{thm} Let $C$ be an additive self-orthogonal subcode of GF$(4)^n$,
containing $2^{n-k}$ vectors, such that there are no vectors of weight
$<d$ in $C\setminus C^\perp$. Then any eigenspace of $\phi^{-1}(C)$ is a
QECC with parameters $[[n,k,d]]$.
\end{thm}

By Glynn (see \cite{Glynn:Graph,Glynn:Tome}), we have: 
Let $S$ be a stabilizer matrix, that is $(n-k)\times n$
over GF(4) and such that its rows are GF(2)-linearly independent.
Then we define a QECC with parameters $[[n,k,d]]$ as the set of all
GF(2)-linear combinations of the rows of $S$. The code is {\em self-dual}
when $k=0$. 

\subsection{The QECC as a Graph}
Assume that each column of $S$ contains at least two non-zero
values, for the columns that do not have this property may be deleted to obtain
a better code.
Following \cite{Glynn:Graph}, a self-dual quantum code
$[[n,0,d]]$ corresponds to a graph on $n$ vertices, which may be assumed to be
connected if the code is indecomposable.
Let PG$(m,q)$ be the finite projective space defined from the vector space of
rank $m+1$ over the field GF$(q)$. Then, the {\em Grassmannian} of lines of
PG$(n-1,2)$, $G_1(\mbox{PG}(n-1,2))$, regarded as a variety immersed in
PG$(\tiny{\left(\begin{array}{c}n\\
2
\end{array}\right)},2)$
is as follows: each line $l_i$ is defined by two points, $a_i$ and $b_i$.
We associate to the set of lines all products
$a_ib_j+a_jb_i,\ i\neq j\ (\mbox{mod} 2)$.
Define a mapping from a column of an $n\times n$ stabilizer matrix $S$
over GF(4) to a vector of length
$\begin{small} {\left(\begin{array}{c}n\\
2
\end{array}\right)} \end{small}$
with coefficients in GF(2): We write each column
over GF(4) as $a+b\omega$, where \linebreak $a,b\in$ GF$(2)^n$.

\begin{small}
$$\left(\begin{array}{c}x_1\\
x_2\\
\vdots\\
x_n
\end{array}\right)=\left(\begin{array}{c}a_1\\
a_2\\
\vdots\\
a_n
\end{array}\right)+\omega\left(\begin{array}{c}b_1\\
b_2\\
\vdots\\
b_n
\end{array}\right) \enspace.$$
\end{small}

Taking all the $2\times 2$ subdeterminants found when we
put the two vectors into a matrix, we get the points of the Grassmannian.
A point in $G_1(\mbox{PG}(n-1,2))\equiv$ a line in PG$(n-1,2)\equiv$ a column of
length $n$ over GF(4) (with at least two different non-zero components).
A quantum self-dual code $[[n,0,d]]$ corresponds to some
set of $n$ lines that generate PG$(n-1,2)$. As each line of PG$(n-1,2)$
corresponds to a (star) kind of graph, the set corresponds to a graph in
$n$ vertices.

\subsection{Interpretation as a Modified Adjacency Generator Matrix over GF(2) and GF(4)}
From any connected graph we obtain an indecomposable code.
Let $\Gamma$ be the adjacency matrix of a graph $G$ in $n$ variables.
Then, $G_T=(I\ |\ \Gamma)$ (where $I$ is the $n\times n$ identity matrix) is
the generator matrix of a binary linear code
\cite{Tonc:Err}. In other words,

\begin{small}
$$G_T=\left(\begin{array}{cccccccc}
1 & 0  &  \ldots & 0 & 0 & a_{01} &  \ldots & a_{0n}\\
0 & 1 &   \ldots & 0 & a_{01} & 0 &  \ldots & a_{1n}\\
\vdots &  \vdots &  \ddots & \vdots & \vdots & \vdots &  \ddots & \vdots\\
0 & 0  &  \ldots & 1 & a_{0n} & a_{1n} &  \ldots & 0\end{array}\right)$$
\end{small}

generates a code over GF(2)$^n$. We can
further interpret $G_T$ as a generating matrix of a code over GF(4)$^n$, as follows \cite{Cald:Qua}:

\begin{small}
$$G=\Gamma+\omega I=\left(\begin{array}{cccc}
\omega & a_{01} &  \ldots & a_{0n}\\
a_{01} & \omega &  \ldots & a_{1n}\\
\vdots & \vdots & \ddots & \vdots\\
a_{0n} & a_{1n} &  \ldots & \omega\end{array}\right)$$
\end{small}

is the generating matrix
of an additive code over GF(4)$^n$. Different graphs may define the same code, but
this relation is 1-1 with respect to LC-equivalence between graphs,
as defined in section \ref{LCGraph}.

\subsection{Interpretation as a Modified Adjacency Matrix over ${\mathbb{Z}}_4$}
Define from a graph with adjacency matrix, $\Gamma$,
the generating matrix of an additive code over ${\mathbb{Z}}_4^n$ as $2\Gamma+I$.
This code has the same weight distribution over ${\mathbb{Z}}_4^n$ as
$\Gamma+\omega I$ over GF$(4)^n$. Once again,
LC-equivalent graphs define equivalent ${\mathbb{Z}}_4$ codes.

\subsection{Interpretation as an Isotropic System}
The graph state can also be viewed as an isotropic system (see \cite{Bou:Iso,Bou:Grph,Bou:Tree,Cou:Ver,MonSar}).

Let $A$ be a 2-dimensional vector space over GF(2). For $x,y \in A$, define a bilinear form,
$<,>$, by
$$<x,y>=\left\{\begin{array}{l}
1\ \m{ if } x\neq y,x\neq0\m{ and }y\neq0\\
0,\m{  otherwise}
\end{array}\right.$$

Let $V$ be a finite set. Define the space of GF(2)-homomorphisms $A^V:V\rightarrow A$.
Define in this GF(2)-vector space a bilinear form as:
$$\m{for }\phi,\psi\in A^V,\ <\phi,\psi>=\sum_{v\in V}<\phi(v),\psi(v)>\ (\m{mod }2) \enspace.$$

\begin{df} Let $L$ be a subspace of $A^V$. Then, $I=(V,L)$ is an {\em isotropic system} if
dim $(L)=|V|$ and $<\phi,\psi>=0\ \forall\ \phi,\psi\in L$.
\end{df}

For a graph $G$, $V(G)$ denotes the set of vertices of $G$.
If $v\in V(G)$, ${\cal{N}}(v)$ denotes the {\em neighbourhood} of vertex $v$,
that is, the set of all its neighbours. For $P\subseteq V$, we set
${\cal{N}}(P)=\sum_{v\in P}{\cal{N}}(v)$. Let $K=\{0,x,y,z\}$  be the Klein group,
which is a 2-dimensional vector space, and set $K'=K\setminus \{0\}$. Note that
$x+y+z=0$.

\begin{lem} (\cite{Bou:Grph}) Let $G$ be a simple graph with vertex set $V$.
Let $\phi,\psi\in K'^V$ such that \linebreak $\phi(v)\neq\psi(v)\ \forall v\in V$, and
set $L=\{\phi(P)+\psi({\cal{N}}(P))\ :\ P\subseteq V\}$. Then $S=(V,L)$ is an
isotropic system.\end{lem}

The triple $\Pi=(G,\phi,\psi)$ is called a {\em graphic presentation} of $S$.

For $\phi\in K^V$, we set $\widehat{\phi}=\{\phi(P)\ :\ P\subseteq V\}$.
$\widehat{\phi}$ is a vector subspace of $K^V$.

\begin{df} For $\psi\in K'^V$, the restricted Tutte-Martin polynomial $m(S,\psi;x)$ is defined by
$$m(I,\psi;x)=\sum(x-1)^{dim(L\cup \widehat{\phi})} \enspace,$$
where the sum is over $\phi\in K'^V$ such that $\phi(v)\neq\psi(v),\ v\in V$.\end{df}

\begin{thm}(\cite{Bou:Grph}) If $G$ is a simple graph and $I$ is the isotropic system
defined by a graphic presentation $(G,\phi,\psi)$, then
$$q(G;x)=m(I,\phi+\psi;x) \enspace,$$
where $q(G;x)$ is the interlace polynomial of $G$.\end{thm}

We mention the interlace polynomial and its relation to our work in
Part II of this paper \cite{RP:BCII}.

\subsection{Interpretation as a Quadratic Boolean Function}
Let
$p({\bf{x}}) : GF(2)^n \rightarrow GF(2)$ be a quadratic boolean function, defined
by its Algebraic Normal Form (ANF),
$ p({\bf{x}}) = \sum_{0\leq i < j\leq n-1} a_{ij}x^ix^j +
\sum_{i =0}^{n-1}b_ix_i+\sum_{i =0}^{n-1}c_i \enspace.$
We associate to $p({\bf{x}})$ the non-directed graph that has
variables as vertices (and vice-versa) \cite{Par:QE}
The adjacency matrix,
$\Gamma$, associated to $p({\bf{x}})$, satisfies $\Gamma(i,j) = \Gamma(j,i) = a_{ij}$, $i < j$,
$\Gamma(i,i) = 0$.

\subsection{Interpretation of a Bipartite Quadratic Boolean Function as a Binary Linear Code}
Quadratic ANFs, as represented by bipartite graphs, have an interpretation as
binary linear codes \cite{Par:QE}: Let ${\bf{T_C}},\ {\bf{T_{C^\perp}}}$ be a bipartite
splitting of $\{0,\ldots,n-1\}$, and let us partition the variable set
${\bf{x}}$ as ${\bf{x}}={\bf{x_C}}\cup{\bf{x_{C^\perp}}}$, where
${\bf{x_C}}=\{x_i: i\in {\bf{T_C}}\}$, and
${\bf{x_{C^\perp}}}=\{x_i: i\in {\bf{T_{C^\perp}}}\}$. Let
$p({\bf{x}})=\sum_k q_k({\bf{x_C}})r_k({\bf{x_{C^\perp}}})$,
 where
 $\mbox{deg}(q_k({\bf{x_C}}))=\mbox{deg}(r_k({\bf{x_{C^\perp}}}))=1\ \forall k$
 (clearly, such a function corresponds to a bipartite graph), and let
 $s({\bf{x}})=(-1)^{p({\bf{x}})}$. Then the action of the transform
 $\bigotimes_{i\in {\bf{T}}}H_i$, with
 ${\bf{T}}={\bf{T_C}}$ or ${\bf{T_{C^\perp}}}$,  on $s({\bf{x}})$ gives
 $s'({\bf{x}})=m({\bf{x}})$, with $m$ the ANF of a Boolean function. $s'$
 is the binary indicator for a binary linear $[n, n-|{\bf{T}}|,d]$
 error correcting code. \begin{footnote}{There is also an equivalent interpretation of
bipartite graphs as {\em{binary matroids}} (e.g. \cite{Cam:Cyc}).
}\end{footnote}

\section{Appendix B - Further Spectral Symmetries of Boolean Functions with respect to
$\{I,H,N\}^n$}
The {\em{power spectrum}} of the WHT
of a boolean function is invariant to within a re-ordering of the spectral elements
after an invertible affine transformation of the variables
of the boolean function
\begin{footnote}{
The {\em{power}} of the $k^{th}$ spectral element, $P_k$, is given by $|P_k|^2$,
where $P_k$ is defined in (\ref{WHT}).
}\end{footnote}.
This implies that bent boolean functions remain
bent after affine transform (see Section \ref{BentSection} for a
discussion of bent properties). However, the set of $\{I,H,N\}^n$ power spectra are
not an invariant of affine transformation.
In this section we ascertain
for which binary transformations (other than LC) the power spectra of the $\{I,H,N\}^n$ transform
remains invariant to within a re-ordering of the spectral elements within each spectrum.
We refer to the complete set of $3^n \times 2^n$ power spectral values w.r.t.
$\{I,H,N\}^n$ as ${\bf{S_{IHN}}}$. Moreover, 'invariance' is to within any re-ordering
of the $3^n \times 2^n$ spectral elements.
From the discussion of sections \ref{LCMultispectra} and \ref{IHNsection}
it is evident that ${\bf{S_{IHN}}}$
of a quadratic boolean function is LC-invariant.
However the LC-orbit is not the only spectral symmetry exhibited with
respect to ${\bf{S_{IHN}}}$. We identify the following symmetries.
\begin{lem}
Let $p({\bf{x}})$ be a boolean function of {\em{any}} degree.
Then ${\bf{S_{IHN}}}$ of $p({\bf{x}})$ and ${\bf{S_{IHN}}}$ of
$p({\bf{x}}) + l({\bf{x}})$ are equivalent, where $l$ is any affine function of
its arguments.
\label{AffOff}
\end{lem}
\begin{lem}
Let $p({\bf{x}})$ be a boolean function of {\em{any}} degree
over $n$ variables.
Then ${\bf{S_{IHN}}}$ of $p({\bf{x}})$ and ${\bf{S_{IHN}}}$ of
$p({\bf{x + a}})$ are
equivalent, where ${\bf{a}} \in {\mbox{GF}}(2)^n$.
\label{CycSym}
\end{lem}
\begin{proof}
Replacing $x_j$ with $x_j + 1$ within any $p({\bf{x}})$
is equivalent to the action of the 'bit-flip' operator,
$\sigma_x = \begin{tiny} \left ( \begin{array}{ll}
0 & 1 \\
1 & 0
\end{array} \right ) \end{tiny}$,
at position $j$ of the transform on $(-1)^{p({\bf{x}})}$, applying $I$ in the rest of the positions.

We can rewrite $H\sigma_x$ as follows,
$$H\sigma_x =
\begin{tiny} \frac{1}{\sqrt{2}}\left ( \begin{array}{ll}
1 & 1 \\
1 & -1
\end{array} \right ) \end{tiny}
\begin{tiny} \left ( \begin{array}{ll}
0 & 1 \\
1 & 0
\end{array} \right ) \end{tiny}
 =
\begin{tiny} \frac{1}{\sqrt{2}} \left ( \begin{array}{ll}
1 & 0 \\
0 & -1
\end{array} \right ) \end{tiny}
\begin{tiny} \left ( \begin{array}{ll}
1 & 1 \\
1 & -1
\end{array} \right ) \end{tiny}
 =
\begin{tiny} \left ( \begin{array}{ll}
1 & 0 \\
0 & -1
\end{array} \right ) \end{tiny}H = \sigma_zH \enspace.
$$
In other words, a bit-flip (or periodic shift)
followed by the action of $H$ is identical to
the action of $H$ followed by a 'phase-flip'. (This is well-known to
quantum code theorists). The final phase-flip is a member of the set ${\bf{D}}$
(see Section \ref{LCLUT} for a definition of ${\bf{D}}$)
so does not change the magnitude
of the spectral values produced by $H$. Therefore the power
spectra produced by $H$ is invariant to prior periodic shift.

We can rewrite $N\sigma_x$ as follows,
$$N\sigma_x =
\begin{tiny} \frac{1}{\sqrt{2}}\left ( \begin{array}{ll}
1 & i \\
1 & -i
\end{array} \right ) \end{tiny}
\begin{tiny} \left ( \begin{array}{ll}
0 & 1 \\
1 & 0
\end{array} \right ) \end{tiny}
 =
\begin{tiny} \frac{1}{\sqrt{2}} \left ( \begin{array}{ll}
0 & i \\
-i & 0
\end{array} \right ) \end{tiny}
\begin{tiny} \left ( \begin{array}{ll}
1 & i \\
1 & -i
\end{array} \right ) \end{tiny}
 =
\begin{tiny} \left ( \begin{array}{ll}
0 & i \\
-i & 0
\end{array} \right ) \end{tiny}N = -\sigma_yN \enspace,
$$
where $\sigma_y$ is one of the four Pauli matrices.
In other words, a bit-flip (or periodic shift)
followed by the action of $N$ is identical to
the action of $N$ followed by a member of the set ${\bf{D}}$.
Therefore the power
spectra produced by $N$ is invariant to a prior periodic shift.

The above argument is trivial with respect to $I$.
The argument extends naturally
to any $n$-dimensional tensor product of $I$, $H$, and $N$.
\end{proof}

Let $p({\bf{x}})$ be a boolean function of {\em{any}} degree
over $n$ variables. We perform a combination of affine offset
and periodic shift on $p({\bf{x}})$ by the following operation:
$$ p({\bf{x}}) \Rightarrow p({\bf{x}} + {\bf{a}}) +
{\bf{c \cdot x}}
 + d \enspace,$$
where ${\bf{a}},{\bf{c}} \in {\mbox{GF}}(2)^n$, $d \in {\mbox{GF}}(2)$,
and '$\cdot$' is the scalar product.

The symmetries generated by affine offset and periodic shift include
all symmetries generated by any combination of
periodic and negaperiodic shift, because we perform periodic and negaperiodic
shifts on $p({\bf{x}})$ by the following operation:
$$ p({\bf{x}}) \Rightarrow p({\bf{x}} + {\bf{a}}) +
{\bf{c \cdot x}}
 + \m{ wt}({\bf{c}}), \mf {\bf{c \preceq a}} \enspace,$$
where ${\bf{a}},{\bf{c}} \in {\mbox{GF}}(2)^n$,
'${\bf{c \preceq a}}$' means that $c_i \le a_i$, $\forall i$ (i.e.
$a$ {\em{covers}} $c$), and $\m{wt}({\bf{c}})$ is the binary
weight of ${\bf{c}}$.
The one positions in ${\bf{a}}$ identify variables $x_i$ which are to
undergo periodic or negaperiodic shift, and the
one positions in ${\bf{c}}$ identify the variables $x_i$ which are to
undergo negaperiodic shift. The combined periodic and negaperiodic symmetry
induced by $\{I,H,N\}^n$ implies an aperiodic symmetry, as discussed further
in \cite{DanAPC}.

\end{document}

%% file: custom1.tex
\newcommand{\beg}{\begin{equation}}
\newcommand{\eeg}{\end{equation}}
\newcommand{\br}{\begin{array}}
\newcommand{\er}{\end{array}}
\newcommand{\bea}{\begin{equation} \begin{array}{c}}
\newcommand{\eea}{\end{array} \end{equation}}
\newcommand{\un}{\underline}
\newcommand{\cb}{\begin{center}}
\newcommand{\ce}{\end{center}}
\newcommand{\llg}{\left\langle}
\newcommand{\rrg}{\right\rangle}
\newcommand{\thh}{$^{\mbox{th}}$}
\newcommand{\al}{\alpha}
\newcommand{\bite}{\begin{itemize}}
\newcommand{\eite}{\end{itemize}}
\newcommand{\m}{\mbox}
\newcommand{\wh}{\hspace{3mm} \mbox{ where }}
\newcommand{\lcm}{\mbox{lcm}}
\newcommand{\di}{\mbox{ div }}
\newcommand{\T}{\mbox{ TRUE }}
\newcommand{\SP}{\mbox{ SPAN}}
\newcommand{\GF}{\mbox{ GF}}
\newcommand{\gf}{\mbox{ {\tiny{GF}}}}
\newcommand{\GR}{\mbox{ GR}}
\newcommand{\Di}{\mbox{ DIV}}
\newcommand{\mt}{\hspace{10mm} \mbox{ }}
\newcommand{\mf}{\hspace{5mm} \mbox{ }}
\newcommand{\mz}{\hspace{2mm} \mbox{ }}
\newcommand{\mv}{\hspace{1mm} \mbox{ }}
\newcommand{\ga}{\gamma}
\newcommand{\xl}{\begin{tiny} \begin{array}{c} < \\ \simeq \end{array} \end{tiny}}
\newcommand{\xm}{\begin{tiny} \begin{array}{c} > \\ \simeq \end{array} \end{tiny}}
\newcommand{\OR}{\mbox{ ord}}
\newcommand{\dg}{\mbox{ deg}}
\newtheorem{thm}{Theorem}
\newtheorem{pro}{Proposition}
\newtheorem{lem}{Lemma}
\newtheorem{cj}{Conjecture}
\newtheorem{cor}{Corollary}
\newtheorem{df}{Definition}
\newtheorem{imp}{Implication}
\newcommand{\mo}{\mbox{ mod }}
\newcommand{\Tr}{\mbox{ Tr}}
\newcommand{\mn}{\mbox{ {\tiny{min}}}}
\newcommand{\mx}{\mbox{ {\tiny{max}}}}
\newcommand{\erf}{\mbox{ erf}}
\newcommand{\erfc}{\mbox{ erfc}}
\newcommand{\SNR}{\mbox{ SNR}}
\newcommand{\BER}{\mbox{ BER}}
\newcommand{\hl}{\\ \hline}
\newcommand{\bd}[1]{\textbf{#1}}

%% file: LCPartIf.bbl
\begin{thebibliography}{99}

\bibitem{Aig:Int}
M. Aigner and H. van der Holst,
{"Interlace Polynomials",}
{\em Linear Algebra and its Applications},
{\bf 377}, pp. 11--30, 2004.

\bibitem{Arr:Int}
R. Arratia, B. Bollobas, and G.B. Sorkin,
{"The Interlace Polynomial: a new graph polynomial",}
{\em Proc. 11th Annual ACM-SIAM Symp. on Discrete Math.},
pp. 237--245, 2000.

\bibitem{Arr:DNA}
R. Arratia, B. Bollobas, D. Coppersmith, and G.B. Sorkin,
{"Euler Circuits and DNA Sequencing by Hybridization",}{\em Disc. App. Math.},
{\bf 104}, pp. 63--96, 2000.

\bibitem{Arr:Int1}
R. Arratia, B. Bollobas, and G.B. Sorkin,
{"The Interlace Polynomial of a Graph",}
{\em J. Combin. Theory Ser. B},
{\bf{92}}, 2, pp. 199--233, 2004.
Preprint:
\href{http://arxiv.org/abs/math/0209045}
{http://arxiv.org/abs/math/0209045},
v2, 13 Aug. 2004.

\bibitem{Arr:Int2}
R. Arratia, B. Bollobas, and G.B. Sorkin,
{"Two-Variable Interlace Polynomial",}
{\em Combinatorica},
{\bf{24}}, 4, pp. 567--584, 2004.
Preprint:
\href{http://arxiv.org/abs/math/0209054}
{http://arxiv.org/abs/math/0209054},
v3, 13 Aug. 2004.

\bibitem{Brie:Ent}
H.J. Briegel and R. Raussendorf,
{"Persistent Entanglement in Arrays of Interacting Particles,"}
{\em quant-ph/0004051 v2},
28 Aug 2000.

\bibitem{Bou:Iso}
A. Bouchet,
{"Isotropic Systems,"}
{\em European J. Combin.},
{\bf 8}, pp. 231--244, 1987.

\bibitem{Bou:Tree}
A. Bouchet,
{"Transforming trees by succesive local complementations"}
{\em J. Graph Theory},
{\bf 12}, pp. 195-207, 1988.

\bibitem{Bou:Grph}
A. Bouchet,
{"Graphic Presentation of Isotropic Systems",}
{\em J. Combin. Thoery B},
{\bf 45}, pp. 58--76, 1988.

\bibitem{Bou:Mart}
A. Bouchet,
{"Tutte-Martin Polynomials and Orienting Vectors of Isotropic Systems",}
{\em Graphs Combin.},
{\bf 7}, pp. 235--252, 1991.

\bibitem{Cald:Qua}
A.R. Calderbank,E.M. Rains,P.W. Shor and N.J.A. Sloane,
{"Quantum Error Correction Via Codes Over $\GF(4)$,"}
{\em IEEE Trans. on Inform. Theory,}
{\bf 44}, pp. 1369--1387, 1998,
(preprint:
\href{http://xxx.soton.ac.uk/abs/quant-ph/?9608006}
{http://xxx.soton.ac.uk/abs/quant-ph/?9608006}).

\bibitem{Cam:Cyc}
P.J.Cameron,
{"Cycle Index, Weight Enumerator, and Tutte Polynomial",}
{\em Electronic Journal of Combinatorics},
{\bf{9}}, 2, 2002.

\bibitem{Car:NewB}
C. Carlet,
{"Two New Classes of Bent Functions"},
{\em Advances in Cryptology - EUROCRYPT'93,
Lecture Notes in Computer Science, Springer-Verlag},
Vol 765, pp. 77--101, 1994.

\bibitem{Cou:Ver}
B. Courcelle and S. Oum,
{"Vertex-minors, MS Logic and Seese's Conjecture"},
{\em preprint}, 2004.

\bibitem{Dan:Dat}
L.E. Danielsen,
{"Database of Self-Dual Quantum Codes"},
\href{http://www.ii.uib.no/~larsed/vncorbits/}
{\it{http://www.ii.uib.no/\~{}larsed/vncorbits/}},
2004.

\bibitem{DanQECC}
L.E. Danielsen,
{\em Master's Thesis - in preparation},
{Selmer Centre, Inst. for Informatics, University
of Bergen, Bergen, Norway}, 2004.

\bibitem{DanAPC}
L.E. Danielsen,T.A. Gulliver and M.G. Parker,
{"Aperiodic Propagation Criteria for Boolean Functions,"}
{\em ECRYPT Document Number: STVL-UiB-1-APC-1.0},
\href{http://www.ii.uib.no/~matthew/GenDiff4.pdf}
{http://www.ii.uib.no/\~{}matthew/GenDiff4.ps},
August 2004.

\bibitem{DanPAR}
L.E. Danielsen and M.G. Parker,
{"Spectral Orbits and Peak-to-Average Power Ratio
of Boolean Functions with respect to the $\{I,H,N\}^n$ Transform",}
{\em SETA'04, Sequences and their Applications, Seoul, Accepted for
Proceedings of SETA04, Lecture Notes in Computer Science, Springer-Verlag, 2005},
\href{http://www.ii.uib.no/~matthew/seta04-parihn.pdf}
{http://www.ii.uib.no/\~{}matthew/seta04-parihn.ps},
October 2004.

\bibitem{Dav:PF}
J.A. Davis and J. Jedwab,
{"Peak-to-mean Power Control in OFDM, Golay Complementary Sequences and Reed-Muller
Codes,"}
{\em IEEE Trans. Inform. Theory},
Vol 45, No 7, pp 2397--2417, Nov 1999.

\bibitem{Dill:DS}
J.F. Dillon,
{"Elementary Hadamard Difference Sets",}
{\em Ph.D. Dissertation, Univ. Maryland, College Park},
1974.

\bibitem{Dob:Bent}
H. Dobbertin,
{"Construction of Bent Functions and Balanced Functions with High Nonlinearity,"}
{\em Fast Software Encryption, Lecture Notes in Computer Science, Springer-Verlag}
No 1008, pp 61-74, 1994.

\bibitem{Glynn:Graph}
D.G. Glynn,
{"On Self-Dual Quantum Codes and Graphs",}
{\em Submitted to the Electronic Journal of Combinatorics},
Preprint at:
\href{http://homepage.mac.com/dglynn/quantum_files/Personal3.html}
{http://homepage.mac.com/dglynn/quantum\_files/Personal3.html},
April 2002.

\bibitem{Glynn:Tome}
D.G. Glynn, T.A. Gulliver, J.G. Maks and M.K. Gupta,
{\em The Geometry of Additive Quantum Codes - Connections with Finite
Geometry,}
Springer-Verlag, 2004.

\bibitem{Gol:Comp}
M.J.E. Golay,
{"Complementary Series",}
{\em IRE Trans. Inform. Theory},
{\bf IT-7}, pp. 82--87, Apr. 1961.

\bibitem{Gras:QG}
M. Grassl,A. Klappenecker and M. Rotteler,
{"Graphs, Quadratic Forms, and Quantum Codes",}
{Proc. IEEE Int. Symp. on Inform. Theory, Lausanne, Switzerland},
June 30-July 5, 2002.

\bibitem{Gras:QECCs}
M. Grassl,
{"Bounds on dmin for additive $[[n,k,d]]$ QECC,"},
\href{http://iaks-www.ira.uka.de/home/grassl/QECC/TableIII.html}
{http://iaks-www.ira.uka.de/home/grassl/QECC/TableIII.html},
Feb. 2003.

\bibitem{Hein:GrEnt}
M. Hein, J. Eisert and H.J. Briegel,
{"Multi-Party Entanglement in Graph States",}
{\em Phys. Rev. A},
{\bf 69}, 6, 2004.
Preprint: \href{http://xxx.soton.ac.uk/abs/quant-ph/0307130}
{http://xxx.soton.ac.uk/abs/quant-ph/0307130}.

\bibitem{Hohn:Klein}
G. Hohn,
{"Self-Dual Codes over the Kleinian Four Group",}
Mathematische Annalen,
{\bf 327}, pp. 227--255, 2003.

\bibitem{Klapp:Cliff1}
A. Klappenecker and M. Rotteler,
{"Clifford Codes"},
Chapter 10, {\bf{Mathematics of Quantum Computation}},
R. Brylinski, G. Chen (eds.), CRC Press, 2002.

\bibitem{MacW:Cod}
F.J.MacWilliams and N.J.A.Sloane,
{\bf The Theory of Error-Correcting Codes},
{Amsterdam: North-Holland},
1977.

\bibitem{Meier:NL}
W. Meier,O. Staffelbach,
{"Nonlinearity Criteria for Cryptographic Functions",}
{\em Advances in Cryptology - EUROCRYPT'89,
Lecture Notes in Computer Science, Springer-Verlag},
Vol 434, pp. 549--562, 1990.

\bibitem{MonSar}
J. Monaghan, I. Sarmiento,
{"Properties of the interlace polynomial via isotropic systems",}
{\em preprint}

\bibitem{Par:Bent}
M.G. Parker,
{"The Constabent Properties of Golay-Davis-Jedwab Sequences",}
{\em Int. Symp. Inform. Theory, Sorrento, Italy},
June 25--30, 2000.

\bibitem{Par:QFG}
M.G. Parker,
{"Quantum Factor Graphs",}
{\em Annals of Telecom.},
July-Aug, pp. 472--483, 2001,
(originally 2nd Int. Symp. on Turbo Codes and Related Topics, Brest, France
Sept 4--7, 2000),
Preprint: \href{http://xxx.soton.ac.uk/ps/quant-ph/0010043}
{http://xxx.soton.ac.uk/ps/quant-ph/0010043}.

\bibitem{Par:QE}
M.G. Parker and V. Rijmen,
{"The Quantum Entanglement of Binary and Bipolar Sequences",}
short version in {\em Sequences and Their Applications},
Discrete Mathematics and
Theoretical Computer Science Series, Springer-Verlag, 2001,
long version at
\href{http://xxx.soton.ac.uk/abs/quant-ph/?0107106}
{http://xxx.soton.ac.uk/abs/quant-ph/?0107106}
or
\href{http://www.ii.uib.no/~matthew/BergDM2.ps}
{http://www.ii.uib.no/\~{}matthew/BergDM2.ps},
June 2001.

\bibitem{Par:LowPAR}
M.G. Parker and C. Tellambura,
"A Construction for Binary Sequence Sets with Low Peak-to-Average Power Ratio",
{\em Technical Report No 242, Dept. of Informatics,
University of Bergen, Norway},
\href{http://www.ii.uib.no/publikasjoner/texrap/ps/2003-242.ps}
{http://www.ii.uib.no/publikasjoner/texrap/ps/2003-242.ps},
Feb 2003.

\bibitem{Par:SB}
M.G. Parker,
"Generalised S-Box Nonlinearity",
{\em NESSIE Public Document - NES/DOC/UIB/WP5/020/A},
\href{https://www.cosic.esat.kuleuven.ac.be/nessie/reports/phase2/SBoxLin.pdf}
{https://www.cosic.esat.kuleuven.ac.be/nessie/reports/phase2/SBoxLin.pdf},
11 Feb, 2003.

\bibitem{Raus:QC}
R. Raussendorf and H.J. Briegel,
{"Quantum Computing via Measurements Only",}
\href{http://xxx.soton.ac.uk/abs/quant-ph/0010033}
{http://xxx.soton.ac.uk/abs/quant-ph/0010033},
7 Oct 2000.

\bibitem{RP:BCII}
C. Riera and M.G. Parker,
{"Generalised Bent Criteria for Boolean Functions (II)",}
\href{http://www.ii.uib.no/~matthew/LCPartIIf.pdf}
{http://www.ii.uib.no/\~{}matthew/LCPartIIf.ps},
2004.

\bibitem{Rud:RS}
W. Rudin,
{"Some Theorems on Fourier Coefficients",}
{\em Proc. Amer. Math. Soc.},
No 10, pp. 855--859, 1959.

\bibitem{Sch:QG}
D. Schlingemann and R.F. Werner,
{"Quantum error-correcting codes associated with graphs"},
{\em Phys. Rev. A},
{\bf 65}, 2002,
\href{http://xxx.soton.ac.uk/abs/quant-ph/?0012111}
{http://xxx.soton.ac.uk/abs/quant-ph/?0012111}, Dec. 2000.

\bibitem{Slo:Seq}
N.J.A. Sloane,
{"The On-Line Encyclopedia of Integer Sequences"},
\href{http://www.research.att.com/~njas/sequences/}
{http://www.research.att.com/\~{}njas/sequences/},
2004.

\bibitem{Tonc:Err}
V.D. Tonchev,
{"Error-correcting codes from graphs"},
{\em Discrete Math.},
Vol. 257, Issues 2--3, 28 Nov., pp. 549--557, 2002.

\bibitem{VanD:Gr}
M. Van den Nest, J. Dehaene and B. De Moor,
{"Graphical description of the action of local {C}lifford
transformations on graph states,"},
{\em Phys. Rev. A},
{\bf 69}, 2, 2004.
Preprint: \href{http://xxx.soton.ac.uk/abs/quant-ph/?0308151}
{http://xxx.soton.ac.uk/abs/quant-ph/?0308151}.

\end{thebibliography}
